\documentclass[structabstract]{aa}

\usepackage{mathrsfs}         
\usepackage{rotating}         
\usepackage{subfigure}        
\usepackage{lscape}           
\usepackage{afterpage}        
\usepackage{xspace}           
\usepackage{natbib}           
\usepackage{graphicx}
\usepackage{epstopdf}
\usepackage{indentfirst}
\usepackage{chemfig}
\usepackage{savesym}
\usepackage{amsmath}
\usepackage{bm}
\savesymbol{iint}
\usepackage{txfonts}  
\restoresymbol{TXF}{iint}
\usepackage{multicol}
\usepackage{booktabs,dcolumn}
\usepackage[T1]{fontenc}
\usepackage{float}
\usepackage{longtable}

\title{The first steps of Interstellar Phosphorus Chemistry
\thanks{Based on observations carried out with the IRAM 30m Telescope. IRAM is supported by INSU/CNRS (France), MPG (Germany), and IGN (Spain).}}

\author{J.~Chantzos\inst{1} \and V. M. ~Rivilla\inst{2}  \and A.~Vasyunin\inst{3,4} \and E.~Redaelli\inst{1}  \and L. Bizzocchi\inst{1} \and F. Fontani\inst{2} \and P.~Caselli\inst{1}}

\institute{Center for Astrochemical Studies, 
           Max-Planck-Institut f\"ur extraterrestrische Physik,
           Gie\ss enbachstra\ss e 1, 85748 Garching (Germany) \and INAF-Osservatorio Astrofisico di Arcetri, Largo Enrico Fermi 5, I-50125, Firenze, Italy \and Ural   
           Federal University, Ekaterinburg, Russia \and Visiting Leading Researcher, Engineering Research Institute
`Ventspils International Radio Astronomy Centre' of Ventspils
University of Applied Sciences, In\v{z}enieru 101, Ventspils LV-3601,
Latvia}

\titlerunning{The first steps of Interstellar Phosphorus Chemistry}
\authorrunning{J. Chantzos et al.}

\begin{document}

\abstract
{Phosphorus-bearing species are an essential key to form life on Earth, however they have barely been detected
in the interstellar medium. Since only PN and PO have been identified so far towards star-forming regions, the chemical formation pathways of P-bearing molecules are not easy to constrain and are thus highly debatable. An important factor still missing in the chemical models is the initial elemental abundance of phosphorus, i.e. the depletion level of P.}
{In order to overcome this problem, we study P-bearing species in diffuse/translucent clouds. In these objects phosphorus is mainly in the gas phase and therefore the elemental initial abundance needed in our chemical simulations corresponds to the cosmic one and is thus well constrained.}
{An advanced chemical model with an updated P-chemistry network has been used. Single-pointing observations were performed with the IRAM 30m telescope towards the line of sight to the blazar B0355+508 aiming for the (2-1) transitions of PN, PO, HCP and CP. This line of sight incorporates five diffuse/translucent clouds.}
{The (2-1) transitions of the PN, PO, HCP and CP were not detected. We report detections of the (1-0) lines of $\mathrm{^{13}CO}$, HNC and CN along with a first detection of $\mathrm{C^{34}S}$. We have reproduced the observations of HNC, CN, CS and CO in every cloud by applying typical conditions for diffuse/translucent clouds.}
{According to our best-fit model, the most abundant P-bearing species are HCP and CP ($\sim10^{-10}$), followed by PN, PO and $\mathrm{PH_3}$ ($\sim10^{-11}$). We show that the production of P-bearing species is favoured towards translucent rather than diffuse clouds, where the environment provides a stronger shielding from the interstellar radiation. Based on our improved model, the (1-0) transitions of HCP, CP, PN and PO are expected to be detectable with estimated intensities up to $\sim200$ mK.}


\maketitle

\section{Introduction} \label{intro}
Phosphorus  has a great relevance for biotic chemistry, since it is a fundamental component of many important biological molecules, such as nucleic acids and phospholipids. P is therefore essential to life on Earth and can consequently play an important role in exoplanets \citep{schaefer}.  
Despite its importance, the chemistry of P-bearing molecules is at its infancy and highly unknown. The aim of this work is to add an important, missing piece to the puzzle: unveiling the first steps of P-chemistry via observations and chemical simulations of simple P-bearing molecules in diffuse clouds.\\  
The ion $\mathrm{P^{+}}$ was detected in several diffuse clouds by \cite{jura}, where an elemental abundance of $\sim2 \times 10^{-7}$ with a low P depletion factor of $\sim2-3$ was derived. However, a more recent study by \cite{lebou}, showed that phosphorus remains mostly undepleted towards diffuse clouds. In addition, P has been identified towards dwarf and giant stars \citep{maas, caffau}, while detections of simple P-bearing molecules (PN, PO, HCP, CP, CCP, NCCP,  $\mathrm{PH_3}$) have been done towards the circumstellar material of carbon- and oxygen-rich stars \citep{agundez07, agundez14a, agundez14b, tenenbaum, DeBeck, ziurys}. The species PN and only very recently PO are the only P-bearing molecules to have been discovered  towards dense star-forming regions \citep{turner, fontani, rivilla16, lefloch, mininni, fontani2019} and molecular clouds in the Galactic Center \citep{rivilla18}. The limited number of observations available makes the chemical pathways for P-bearing chemistry strongly debatable. The main uncertainty is the unknown depletion factor of P in molecular clouds. In general, chemical models of dark clouds start with the so called "low-metal abundances", where the elemental abundances of heavy elements (such as P, S, Fe, Mg) are reduced by orders of magnitude to reproduce molecular observations \citep[e.g.][] {agundez13}, but with poor understanding of the chemical processes at the base of such depletions. In the case of P, the level of depletion is still very uncertain. While \cite{turner1990} and \cite{wakelam} used high depletion factors of 600-$10^4$  with respect to P cosmic abundance, recent works have shown that it could be as low as $\sim100$ \citep{rivilla16, lefloch}. 

As only a very limited number of P-bearing molecules have been detected in star-forming regions, it is very hard to put constraints on the elemental abundance of P in the gas phase and on the major chemical pathways. 
In order to elucidate the interstellar P-chemistry we aim at focusing on diffuse clouds, which represent the first steps of molecular-cloud evolution.  Diffuse clouds can give us important constraints on the P-chemistry, since in these objects P is not strongly affected by depletion so that the initial P-abundance that can be used for chemical simulations is well constrained \citep{lebou}.  With this approach we are able to remove an important uncertainty in our model and use a reliable starting point for our chemical simulations. 

So far there have been several chemical and physical models focusing solely on diffuse clouds \citep[e.g.][] {dalgarno88, lepetit04, cecchi12, godard14}. For example, in \cite{godard14}, a model including dissipation of turbulence was applied to reproduce the observed molecular abundances in the diffuse interstellar medium (ISM). The main results showed that chemical complexity is strongly linked to turbulent dissipation, which was able to reproduce the high abundances of CO and other species (such as $\mathrm{C^+}$ and $\mathrm{HCO^+}$) observed towards Galactic diffuse clouds. 
 \cite{lepetit04} describe the development of a chemical model of the diffuse cloud towards $\zeta$ Persei, that was able to reproduce the abundance of $\mathrm{H_3^+}$ and other species, like CN and CO. This was achieved by modelling two phases, namely a small dense phase ($\sim 100 \, \mathrm{au}$) with a density of $n\mathrm{(H)} = 2\times10^4 \, \mathrm{cm^{-3}}$ and a larger diffuse region (4 pc) with $n\mathrm{(H)} = 100 \, \mathrm{cm^{-3}}$. In addition, the reproduction of the  $\mathrm{CH^+}$ abundance and that of the rotationally excited $\mathrm{H_2}$ required the inclusion of shocks into the model. Similar results were achieved by \cite{cecchi12} when including the injection of hot $\mathrm{H_2}$ into the model.  

Previous observations \citep{corby, liszt18, thiel} prove the chemical complexity and the wide range of densities, temperatures and visual extinctions of diffuse and translucent clouds, making them promising targets for observations of P-bearing molecules. Diffuse clouds are characterised by low densities with $n\mathrm{(H)} = 100-500 \, \mathrm{cm^{-3}}$ and are thus more exposed to the interstellar radiation, which can destroy molecules. Translucent clouds on the other hand, are an intermediate state between diffuse and dense molecular clouds, being more protected by UV radiation ($1 \, \mathrm{mag} < A_V < 5 \, \mathrm{mag}$). They are denser with typical densities of $n\mathrm{(H)} = 500-5000 \, \mathrm{cm^{-3}}$ and subsequently cooler ($T_{\mathrm{gas}}= 15-50 \, \mathrm{K}$), showing higher chemical complexity \citep{snow, thiel}. 
One prominent candidate that has been widely studied in previous works \citep[e.g.][and references therein]{liszt18} is the gas that lies along the line of sight to the compact extragalactic continuum source B0355+508. This strong blazar is located at a very low latitude in the outer Galaxy ($b= -1.6037^\circ$), meaning that the way through the Galactic disk is long and therefore gathers a significant amount of distributed Galactic diffuse gas \citep{pety}. In fact, the line of sight towards B0355+508 shows a complex kinematic structure which incorporates several diffuse/translucent clouds. The detections of numerous molecules like sulphur- and CN-bearing species as well as small hydrocarbons towards B0355+508  also indicate the rich chemistry present in this diffuse and translucent gas  \citep[e.g.][and references therein]{liszt18}.
The substantial velocity structure coupled with a high chemical complexity of this line of sight enables us to adjust our chemical/physical model to every cloud component and find which physical conditions favour the abundances of P-bearing molecules the most. Other background sources that have been previously studied are either lacking the chemical (like B0224+671) or the velocity (such as B0415+479) features which are essential for the present work. 

In this paper, we present single-pointing observations of the (2-1) transitions of HCP, CP, PN and PO and chemical simulations of the their molecular abundances towards the line of sight to B0355+508 in order to investigate P-bearing chemistry within diffuse/translucent clouds, the precursors of molecular clouds. In Section \ref{observations} we describe the observational details. Section \ref{results} summarizes the results of the observations.
In Section \ref{model} we describe our updated phosphorus chemical network as well as the grid of models that we apply in order to reproduce the observations of HNC, CN, CS and CO towards every cloud component along the line of sight. Furthermore, in Section \ref{P-stuff} we focus on the P-bearing chemistry based on our best-fit model (that was determined in Section \ref{model}). In particular, we report the predicted molecular abundances of HCP, CP, PN, PO and $\mathrm{PH_3}$ and we study their dependence on the visual extinction, the cosmic-ray ionisation rate as well as the diffusion/desorption energy ratio on dust grains.  The outlook and conclusions are summarized in Sections \ref{future} and  \ref{outlook}.

\section{Observations} \label{observations}
The observations of the HCP (2-1), CP (2-1), PN (2-1) and PO (2-1) transitions in the 3 mm range were carried out at the IRAM 30m telescope located at Pico Veleta (Spain) towards the line of sight to the compact extragalactic quasar B0355+508. Table \ref{tab:observed species} lists the observed transitions, the spectroscopic constants as well as the telescope settings at the targeted frequencies:  the upper state energy is described by $E_{\mathrm{up}}$, the upper state degeneracy is given by $g_u$, while $A_{ul}$ stands for the Einstein coefficient of the transition $u \rightarrow l$. The main beam efficiency and the main beam size of the telescope at a given frequency are denoted by the parameters $B_{\mathrm{eff}}$ and $\theta_{\mathrm{MB}}$, respectively. For our observations we used the EMIR receiver with the E090 configuration (3 mm atmospheric window). We applied three observational set ups, in which every set up covered a total spectral coverage of 7.2 GHz (each sub-band covered 1.8 GHz). As a backend we used the Fast Fourier Transform Spectrometer with a frequency resolution of 50 kHz ($\mathrm{0.15 \, \mathrm{km \, s^{-1}}}$ at 100 GHz).  In addition, we applied the wobbler switching mode with an amplitude offset of $\pm90''$. Pointing  and focus of the telescope was performed every 2 hr on the background source B0355+508 itself and was found to be accurate within 2$''$.

\begin{table*} [! h]
\center
\caption{Spectroscopic parameters of the observed species and telescope settings.}
\label{tab:observed species}
\setlength{\tabcolsep}{10pt}
\begin{tabular} {c c c c c c c c c} 
\hline \hline \\
Species & Transitions & $E_{\mathrm{up}}$ & Frequency & $A_\mathrm{ul}$ & $g_u$  & $B_{eff}$ & $\theta_{\mathrm{MB}}$ & References \\
& & (K) & (GHz) & ($\mathrm{10^{-5} \, s^{-1}}$) & & ($\%$) & ($\arcsec$) & \\
\hline \\ [-1ex ] 
HCP & J=2-1 & 5.8 & 79.90329 &  0.04 & 5 & 83 & 31 & 1  \\

PN & J=2-1 & 6.8 & 93.97977 &  2.92 & 5 &  80 & 26 &  2  \\

CP  &  N =2-1, J=3/2-1/2, F = 2-1 & 6.8 & 95.16416 & 0.33 & 5 & 80 & 26 &  3 \\

PO & J=5/2-3/2, $\Omega=1/2$, F=3-2, e & 8.4  & 108.99845 &  2.13 & 7 & 78  & 23 & 4 \\

PO & J=5/2-3/2, $\Omega=1/2$, F=2-1, e &  8.4 & 109.04540 & 1.92 & 5 & 78 & 23 & 4 \\

PO &  J=5/2-3/2, $\Omega=1/2$, F=3-2, f &  8.4 & 109.20620 & 2.14 & 7 & 78 & 23 & 4 \\

PO & J=5/2-3/2, $\Omega=1/2$, F=2-1, f & 8.4 & 109.28119 & 1.93 & 5 & 78 & 23 &  4 \\
\hline
\end{tabular} 
\tablebib{(1) \cite{bizzocchi}; (2) \cite{cazzoli}; (3) \cite{saito}; (4) \cite{bailleux}.}
\end{table*}

The intensity of the obtained spectra was converted from antenna ($T^{*}_{\mathrm{A}}$) to main beam temperature ($T_{\mathrm{mb}}$) units, using the following relation:  $T_{\mathrm{mb}} = \frac{F_{\mathrm{eff}}}{B_{\mathrm{eff}}} \times {T^{*}_{\mathrm{A}}}$, where $F_{\mathrm{eff}}$ is the forward efficiency. $F_{\mathrm{eff}}$ is equal to 95$\%$  in the targeted frequency range.

\section{Results} \label{results}

The compact extragalactic source B0355+508 is located at $\mathrm{\alpha = 3h \, 59m \, 29.73s}$, $\mathrm{\delta = 50^\circ 57' 50.2''}$ with a low galactic latitude of $b= -1.6037^\circ$, incorporating a large amount of Galactic gas along the line of sight that harbors up to five diffuse/translucent clouds at the velocities of $ -4,\, -8, \, -10, \, -14 \, \mathrm{and} \, -17 \, \mathrm{km \, s^{-1}}$  \citep[e.g.][and references therein]{liszt18}.
The flux of the blazar B0355+508 is variable over time and has been measured at $\sim3$ mm to be on average equal to $(4.62 \pm 1.02)$ Jy, after averaging the flux of 76 different observations \citep{agudo}. This corresponds to a temperature $T_{\mathrm{c}}$ of $(0.96 \pm 0.21)$ K at a beam size of 27$''$ by taking into account the Rayleigh-Jeans-Approximation. 
The obtained spectra were reduced and analyzed by the GILDAS software \citep{pety_gildas}. Every detected line was fitted via the standard CLASS Gaussian fitting method.
For the derivation of the peak opacity we use the radiative transfer equation

\begin{eqnarray} \label{Eq:opacity}
T_{\mathrm{mb}} &=& (J_{\nu}(T_{\mathrm{ex}})-J_{\nu}(T_{\mathrm{bg}}) - J_{\nu}(T_{\mathrm{c}}))\times(1-\exp{(-\tau)})   \Rightarrow \nonumber \\ 
  \tau &=& -\ln \Bigl(1 - \frac{T_{\mathrm{mb}}}{J_{\nu}(T_{\mathrm{ex}})-J_{\nu}(T_{\mathrm{bg}})-J_{\nu}(T_{\mathrm{c}})}\Bigr),
\end{eqnarray} 

\noindent{where $T_{\mathrm{ex}}$ is the excitation temperature, $T_{\mathrm{bg}}$ is the cosmic background temperature and  $J(T) = (\frac{h \nu}{k_B})(e^{\frac{\mathrm{h \nu}}{k_{B} T }}-1)^{-1}$ describes the Rayleigh-Jeans temperature in K\footnote{In case of an emission line, $J_{\nu}(T_{\mathrm{c}})$ is neglected because $J_{\nu}(T_{\mathrm{ex}}) \gg J_{\nu}(T_{\mathrm{c}})$.}.} After obtaining the peak opacity $\tau$, the column density $N$ is then estimated by following the relation:

\begin{eqnarray} \label{Eq:column_density}
N = \tau \sqrt{\frac{16\pi^3}{\ln2}} \frac{\nu^3 Q_{\mathrm{rot}}(T_{\mathrm{ex}}) \Delta \varv \, e^{{E_u / k_B T_{\mathrm{ex}}}}}{c^3 A_{ul} \, g_u \, (e^{h \nu / k_B  T_{\mathrm{ex}}}-1)},
\end{eqnarray}

\noindent{with $k_B$ being the Boltzmann constant, $\Delta \varv$ is the linewidth (FWHM), $\nu$ is the transition frequency, $c$ is the speed of light, and $h$ stands for the Planck constant. $Q_{\mathrm{rot}}(T_{\mathrm{ex}})$  gives the partition function of a molecule at a given excitation temperature $T_{\mathrm{ex}}$.}

\begin{figure*}[h!]
	\centering
 	\includegraphics[width = 0.7\textwidth]{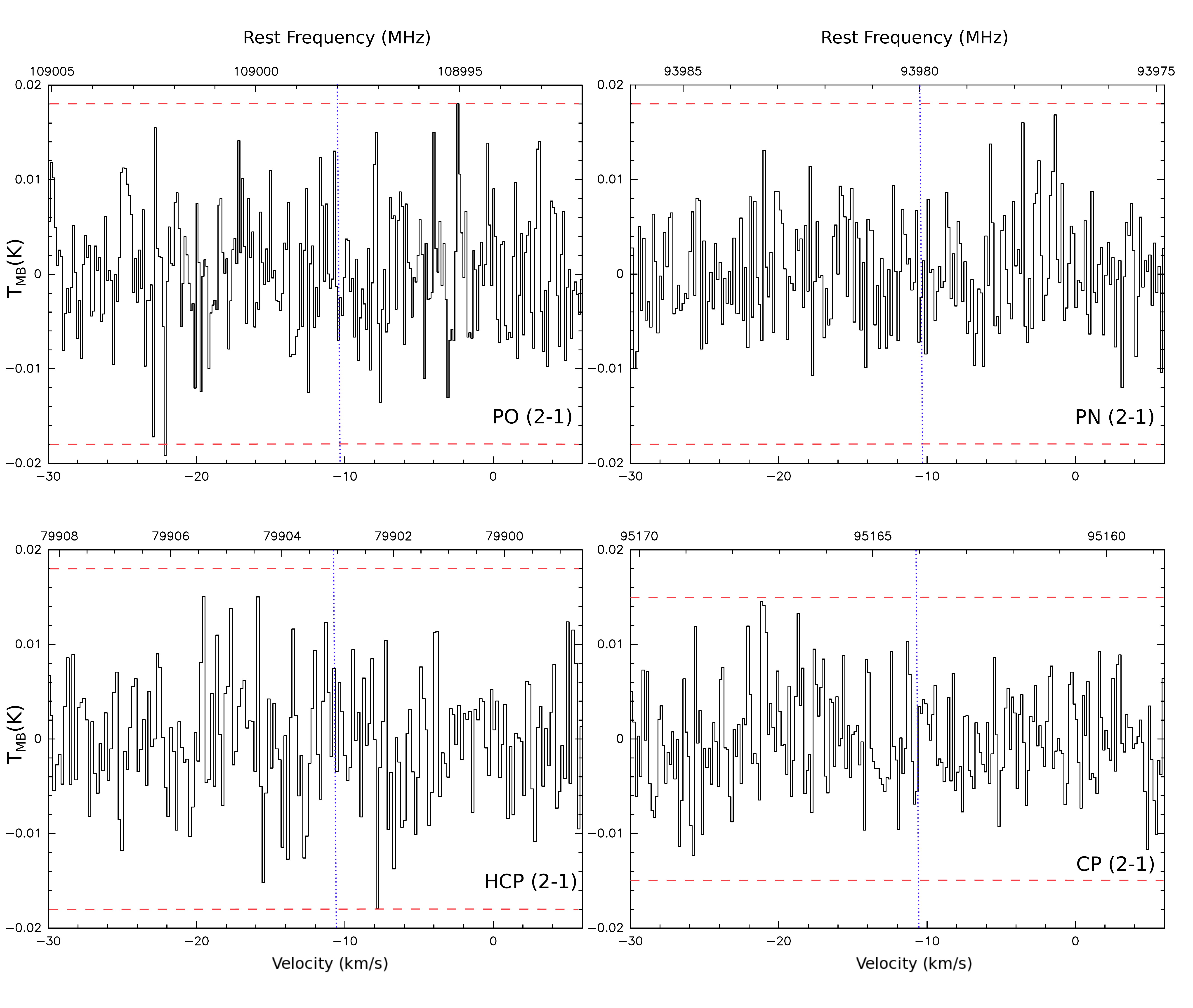}
	\caption{Spectra of the non-detected (2-1) transitions of PO, PN, HCP and CP. The upper x-axis shows the rest frequency  (in MHz) and the lower one is a velocity axis (in  $\mathrm{km \, s^{-1}}$). The red dashed line indicates the $3\sigma$ level and the blue dashed line shows the transition frequency of the corresponding molecule. In case of PO, we show  as an example one of the  observed transitions at 108.998 GHz.}
	\label{Fig:PN_PO_HCP_CP}
\end{figure*}

The (2-1) transitions of HCP, CP, PN and PO were not detected within our observations (see Figure \ref{Fig:PN_PO_HCP_CP}). We derive 3$\sigma$ upper limits for the opacities and column densities of the P-bearing species by using Eqs. \eqref{Eq:opacity} and \eqref{Eq:column_density}.  Due to the low densities in diffuse clouds, molecules are expected to show no collisional excitation. Thus, the column densities were calculated assuming $T_{\mathrm{ex}} = T_{\mathrm{bg}} = 2.7 \, \mathrm{K}$, which simplifies Eq. \eqref{Eq:opacity} to: 

\begin{eqnarray} \label{Eq:opacity_simplified}
  \tau &=& -\ln \Bigl(1 + \frac{T_{\mathrm{mb}}}{J_{\nu}(T_{\mathrm{c}})}\Bigr).
\end{eqnarray} 
\noindent{The results are summarized in Table \ref{tab:upper_limits_PN_PO}.}

\begin{table*} [h]
\centering
\caption{Derived upper limits for the opacity and the column density of HCP, CP, PN and PO.}
\label{tab:upper_limits_PN_PO}
\setlength{\tabcolsep}{10pt}
\begin{tabular} {c c c c c c } 
\hline \hline \\
Species & Frequency  &  $\tau$ & N  & $\mathrm{T_{MB}}$ & rms \\
  & (GHz)& &  $(10^{11} \, \mathrm{cm^{-2}})$ & (K) & (mK) \\
\hline \\ [-1ex]
HCP & 79.90329 & $<0.02$ & $<22.7$ & $<0.02$ & 6  \\
CP & 95.16416 & $<0.02$ & $<12.6$ & $<0.02$  & 5  \\
PN & 93.97977  & $<0.02$ & $<0.42$ & $<0.02$ & 6  \\
PO &  108.99845 & $<0.02$ & $<4.29$ & $<0.02$  & 6  \\
      &  109.04540 & $<0.02$ &$<6.70$ & $<0.02$   & 6  \\
      &  109.20620 & $<0.02$ & $<4.34$ &  $<0.02$ & 6 \\
      &  109.28119 &  $<0.02$ & $<6.69$ & $<0.02$ & 6 \\
\hline \\
\end{tabular} \\
\tablefoot{The upper limits are 3$\mathrm{\sigma}$.}
\end{table*}

\begin{figure}[h]
	\centering
 	\includegraphics[width = 0.4\textwidth]{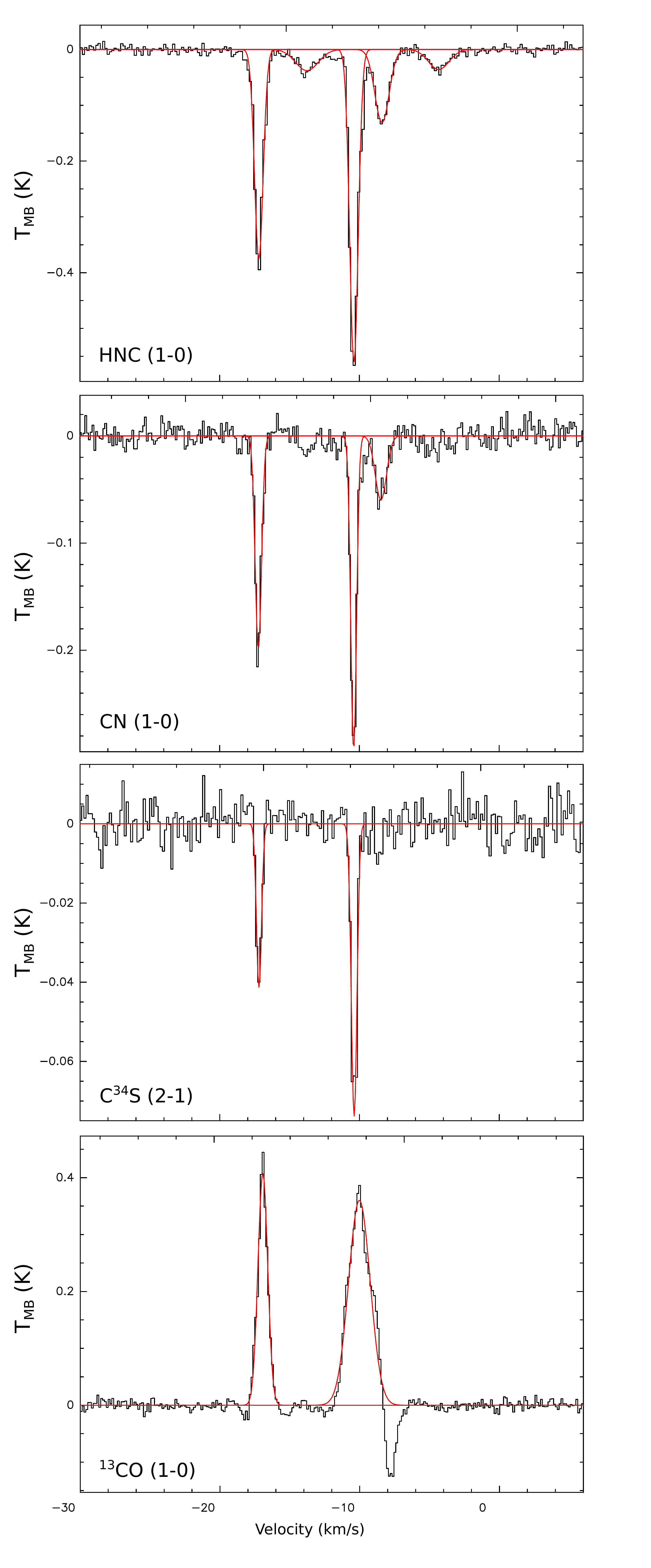}
	\caption{Spectra of the detected species HNC, CN, $\mathrm{C^{34}S}$ and$\mathrm{^{13}CO}$ in the 3 mm range towards the line of sight to the extragalactitc source B0355+508. The red line represents the CLASS Gaussian fit.}
	\label{Fig:all_spectra}
\end{figure}

\begin{figure*}[h]
	\centering
 	\includegraphics[width = 1.0\textwidth]{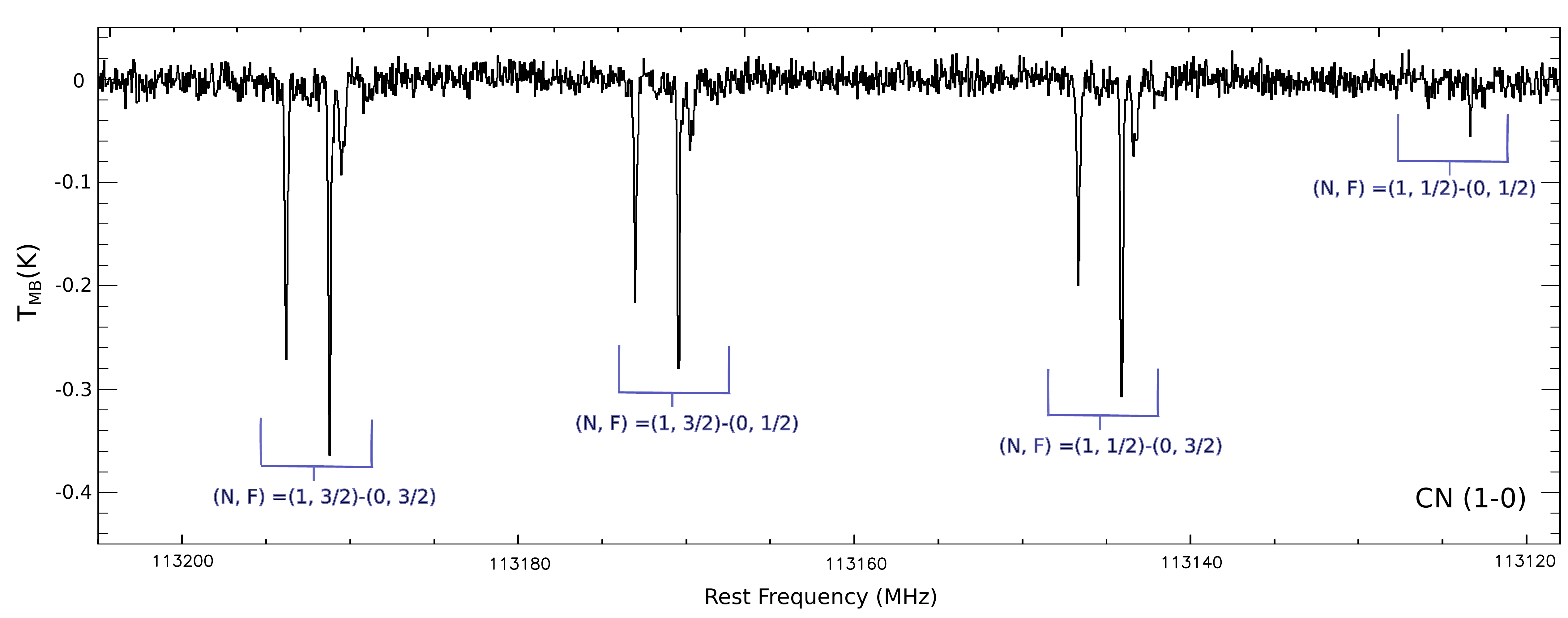}
	\caption{Detected hyperfine components of the CN (1-0) transition between 113.12 and 113.20 GHz. The three strongest hyperfine components were detected in the three clouds with $\varv_{\mathrm{LSR}}=-8, \, -10, \, -17 \,\mathrm{km \, s^{-1}}$ except for the one weak transition $\mathrm{(N, F) = (1, 1/2) - (0, 1/2)}$, which was identified only in the two densest clouds  (at $-10, \, -17 \, \mathrm{km \, s^{-1}}$). }
	\label{Fig:spectrum}
\end{figure*}

We have detected the HNC (1-0), CN (1-0) and $\mathrm{C^{34}S}$ (2-1) transitions in absorption as well as the $\mathrm{^{13}CO}$ (1-0) in emission at the 3 mm range with a high signal-to-noise-ratio (S/N), ranging from 6 to 80 \footnote{The rms levels are lying between 4 and 13 mK.}. Figure \ref{Fig:all_spectra} shows all the detected spectra towards the line of sight to B0355+508. In case of CN we were able to detect and resolve four hyperfine components from 113.123 GHz to 113.191 GHz (see Figure \ref{Fig:spectrum}). Every hyperfine component was detected in the three velocity components at $-8, \, -10, \, -17 \,\mathrm{km \, s^{-1}}$ except for the one weak transition at 113.123 GHz, which was identified only in two clouds  (at $-10, \, -17 \, \mathrm{km \, s^{-1}}$).  The molecule HNC was identified in all five cloud components, while $\mathrm{C^{34}S}$ (in absorption) and $\mathrm{^{13}CO}$ (in emission) were detected solely towards the densest features, at $-10 \, \mathrm{km \, s^{-1}}$ and $-17  \, \mathrm{km \, s^{-1}}$. 
Table \ref{tab:detected species} lists the identified species and the corresponding spectroscopic parameters.

\begin{table*} [! h]
\center
\caption{Spectroscopic parameters of the detected species and telescope settings.}
\label{tab:detected species}
\setlength{\tabcolsep}{10pt}
\begin{tabular}{c c c c c c c c c}
\hline \hline \\
Species & Transitions & $E_{\mathrm{up}}$  & Frequency & $A_\mathrm{ul}$ & $g_u$  & $B_{eff}$ & $\theta_{\mathrm{MB}}$  & References \\
 & & (K) & (GHz)  & $\mathrm{10^{-5} \, s^{-1}}$ & & ($\%$) & ($\arcsec$) &\\
\hline \\ [-1ex ] 
HNC & J=1-0 &  4.4  & 90.66357 &  2.69 & 3 & 81 &  27 & 1  \\

CN & N= 1-0,  F = 3/2-1/2 & 5.5 & 113.17049 &  0.51 & 4 & 78 & 22 &  2 \\

$\mathrm{C^{34}S}$  & J= 2-1 &  6.9 & 96.41295 & 1.60 & 5 & 80 & 26 & 3 \\

$\mathrm{^{13}CO}$ & J=1-0 & 5.3  & 110.20135 & 0.006 & 6 & 78 &  22 &  4\\
\hline
\end{tabular} 
\tablebib{(1) \cite{saykally}; (2) \cite{dixon}; (3) \cite{gottlieb}; (4) \cite{klapper, cazzoli04}.}
\end{table*}

For estimating the CN column density we use the hyperfine component at 113.170 GHz. 
Our derived opacities and column densities of CN agree within a factor of 2-3 with previous results  \citep{liszt01}, while the HNC results are well reproduced within a factor of 1.5. Table \ref{tab:CN} summarizes the derived opacities and column densities of the detected species, as well as the obtained line intensities and rms levels. 

\begin{table*} [h]
\centering
\caption{Gaussian fitting results of CN, HNC, $\mathrm{C^{34}S}$, $\mathrm{^{13}CO}$.}
\label{tab:CN}
\tabcolsep 2.8pt
\begin{tabular} {c c c c c c c c c} 
\hline \hline \\
Species & Velocity  & $\Delta \varv$& $\tau$ &  $\mathrm{T_{MB}}$ & rms & Spectral & N  & N  \\
  & $(\mathrm{km \, s^{-1}})$ & $(\mathrm{km \, s^{-1}})$ & &  (K) & (mK) &  resolution & $(\mathrm{cm^{-2}})$   & $(\mathrm{cm^{-2}})$    \\
   &  & & &  &  & $(\mathrm{km \, s^{-1}})$ &  (this work) &  (previous work)$^{(1)}$ \\
\hline \\ [-1ex]
CN &  &  &  &  & &\\
& $-17.23 \pm 0.01$& $0.54 \pm 0.02$ & $0.23\pm0.07$ &  $0.20\pm0.02$ & 9 & 0.13 &$(0.87\pm0.28) \times 10^{13}$ & $(2.14\pm0.24) \times 10^{13}$  \\
& $-10.41 \pm 0.01$  & $0.49 \pm 0.01$  & $0.36 \pm 0.09$ & $0.29\pm0.02$ & 9 & 0.13 & $(1.23\pm0.34) \times 10^{13}$ & ($3.34\pm0.37)\times 10^{13}$   \\
& $-8.47\pm0.04$ & $0.97\pm0.09$ & $0.06\pm0.03$ &  $0.06\pm0.01$ & 9 & 0.13 & $(0.43\pm0.20) \times 10^{13}$ & $(0.76\pm0.21)\times 10^{13}$  \\
\hline \\
HNC &  &  & & & &\\
 & $-17.197\pm0.004$ &  $0.73\pm0.01$ & $0.50\pm0.10$ &  $0.37\pm0.01$ & 6 & 0.17 & $(0.69\pm0.16)\times 10^{12}$ & $(0.74\pm0.02)\times 10^{12}$   \\
 & $-10.392\pm0.003$ & $0.73\pm0.01$ & $0.87\pm0.16$ &  $0.56\pm0.01$ & 7 & 0.17 &  $(1.20\pm0.23)\times 10^{12}$ & $(1.14\pm0.04)\times 10^{12}$  \\
 & $-8.37\pm0.01$ & $1.14\pm0.04$  &  $0.15\pm0.04$ & $0.13\pm0.01$ & 6 & 0.17 & $(0.32\pm0.15)\times 10^{12}$ & $(0.39\pm0.02)\times 10^{12}$   \\
& $-13.68\pm0.07$ & $2.30\pm0.23$ & $0.04\pm0.01$ & $0.04\pm0.01$  & 6 & 0.17 &  $(0.16\pm0.08)\times 10^{12}$  & $(0.10\pm0.02)\times 10^{12}$ \\ 
 & $-4.32\pm0.06$ & $1.71\pm0.15$ & $0.04\pm0.02$ &  $0.04\pm0.01$ & 6 & 0.17 & $(0.12\pm0.06)\times 10^{12}$ & $(0.14\pm0.01)\times 10^{12}$  \\
 \hline \\
 $\mathrm{C^{34}S}$ &  &  & & &  & \\
 &$-17.20\pm0.02$ &  $0.43\pm0.04$ & $0.04\pm0.02$  & $0.04\pm0.01$ & 4  & 0.16 &  $(1.64\pm0.82)\times 10^{11}$ & $<3.2\times 10^{11}$  $^{(2)}$  \\
 & $-10.39\pm0.01$  & $0.48\pm0.02$  & $0.08\pm0.03$ & $0.07\pm0.01$ & 4 & 0.16 & $(3.33\pm1.23)\times 10^{11}$ &  $<4.4\times 10^{11}$   $^{(2)}$\\
 \hline \\
$\mathrm{^{13}CO}$ &  &  & & & &  \\
 & $-16.935\pm0.004$ &  $0.82\pm0.01$  & $0.154\pm0.004$ &  $0.41\pm0.01$ & 8 & 0.14 & $(3.98\pm0.16)\times 10^{14}$ & $(4.34\pm0.51)\times 10^{14}$ $^{(3)}$   \\
& $-10.03\pm0.01$ &  $1.82\pm0.03$ &$0.135\pm0.004$ & $0.36\pm0.01$ & 13 & 0.14 &  $(7.73\pm0.38)\times 10^{14}$ & $(1.79\pm0.26)\times 10^{14}$  $^{(3)}$\\
\hline \\
\end{tabular} \\
\tablebib{(1) \cite{liszt01}, (2) \cite{liszt_isotope}, (3) \cite{liszt98}.}
\end{table*}

\cite{liszt02} reported the detection of the main isotopologue $\mathrm{C^{32}S}$ done with the IRAM Plateau de Bure interferometer (PdBI) and estimated a column density of $(4.27 \pm 0.16) \times 10^{12} \, \mathrm{cm^{-2}}$ at the $-10 \,\mathrm{km s^{-1}}$ component and  $(3.06 \pm 0.32) \times 10^{12} \, \mathrm{cm^{-2}}$ at  $-17 \, \mathrm{km s^{-1}}$. With the above values and the column densities of $\mathrm{C^{34}S}$ calculated in this work, we derive a sulfur isotopic ratio $\mathrm{^{32}S/^{34}S}$ ratio of $12.8\pm4.8$ and $18.7\pm9.5$  for the components at $-10 \, \mathrm{km s^{-1}}$ and  $-17 \, \mathrm{km s^{-1}}$, respectively. The latter value is in good agreement with the $\mathrm{^{32}S/^{34}S}$ ratio for the local ISM of $24 \pm 5$  \citep{chin1996}. However, the isotopic ratio determined for $\varv = -10 \, \mathrm{km s^{-1}}$ is quite lower than the local interstellar value, which could be the result of opacity effects of the $\mathrm{C^{32}S}$ line. In addition, the determination of the sulphur isotopic ratio was based on just one spectral line of the main species $\mathrm{C^{32}S}$ and its isotopologue,  which also yields a high uncertainty. To our knowledge this is the first detection of $\mathrm{C^{34}S}$ towards this line of sight, owing to the high spectral resolution of $\sim50$ kHz and high sensitivity (rms of $\sim4$ mK) achieved with our observations. 

In \cite{liszt98}, detections of the main species $\mathrm{^{12}CO}$ and its isotopologue $\mathrm{^{13}CO}$ are reported, which were obtained with the PdBI as well as the NRAO 12m telescope. The single-dish observations covered a large beam of 60\arcsec, thus seeing CO and its isotopologue in emission, while the interferometric observations were sensitive only to the very narrow column of gas towards the strong background blazar, giving rise to absorption lines. 
For deriving the excitation temperatures and the column densities, both emission (single-dish data) and absorption lines (interferometric data) were considered. 

In Figure \ref{Fig:all_spectra} it is clearly visible that the strong $\mathrm{^{13}CO}$  emission line at $-10 \,\mathrm{km s^{-1}}$ overlaps with an absorption feature at around $-8 \,\mathrm{km s^{-1}}$. This is probably due to the fact that absorption is present close to the background source, so that emission and absorption lines are merged together in our observations with the IRAM 30m telescope. This contamination effect is influencing the line profile at $-10 \,\mathrm{km s^{-1}}$ which subsequently results in an unreliable fit. This could possibly explain why the $\mathrm{^{13}CO}$  column density derived at $-10 \,\mathrm{km s^{-1}}$ deviates by a factor of $\sim4$ from previous results  \citep{liszt98}, while towards $-17 \,\mathrm{km s^{-1}}$, $N(\mathrm{^{13}CO})$ is well reproduced within 10\% (see Table \ref{tab:CN})\footnote{We used a $T_{\mathrm{ex}}$ of 6 K for deriving $N(\mathrm{^{13}CO})$, as it was inferred in \cite{liszt98}.}. 
Our derived isotopic ratio $\mathrm{^{12}CO/^{13}CO}$ at  $-17 \, \mathrm{km s^{-1}}$ is equal to $16.7\pm1.4$. Herefore we used the column density of $\mathrm{^{12}CO}$ derived in \cite{liszt98} with $N(\mathrm{^{12}CO}) = \mathrm{(6.64 \pm 0.47) \times 10^{15}} \, \mathrm{cm^{-2}}$. The resulting CO isotopic ratio is almost a factor $\sim4$ lower than the local interstellar ratio $\mathrm{^{12}C/^{13}C}= 60$ \citep{liszt_isotope}. This was already confirmed by previous studies \citep{liszt07, liszt17} that show an increased insertion of $\mathrm{^{13}C}$ into CO towards clouds in the translucent regime with elevated densities and/or smaller radiation fields, which lead to an enhanced abundance of $\mathrm{^{13}CO}$ by a factor of 2-4. Under these conditions isotope exchange fractionation ($\mathrm{^{13}C^+} + \mathrm{^{12}CO} \rightarrow \mathrm{^{12}C^{+}} + \mathrm{^{13}CO + 35 \, K}$) is more dominant than selective photodissociation.

\section{Chemical Modeling} \label{model}
The goal of the present study is to constrain and improve our model of diffuse and translucent clouds to make reliable predictions for the abundances of phosphorus-bearing species (and also others). For this reason, we have used the observations of HNC, CN, CS and CO in order to constrain the physical parameters in our model.
The chemical code that we have applied was developed by \cite{vasyunin2013} with an updated grain-surface chemistry (Vasyunin et al. 2019, in prep.). The model includes  a gas-grain chemical network with 6000 gas-phase reactions, 200 surface reactions and 660 species.  Accretion and desorption processes regulate and connect the gas-phase and grain surface chemistry.  The code numerically solves coupled differential equations (chemical rate equations) and computes a set of time dependent molecular abundances. 
Since  the observations were carried out towards diffuse/translucent clouds, we have considered as initial elemental abundances the standard Solar elemental composition as reported in \cite{asplund} (see Table \ref{tab:initial}). We note that our initial elemental abundances are significantly different compared to the low metal abundances used in \cite{wakelam} for dark clouds (200 times more abundant S and up to  $10^4$ more abundant Fe, Cl, P, and F). In particular, the initial abundance of P is $2.6\times10^{-7}$ and thus well constrained unlike in dense molecular clouds.  This approach will help us elucidate much better the chemistry of P since a key parameter for the chemical model is well determined. In addition, we begin our chemical simulations with hydrogen being completely in its atomic form, in order to have pure atomic diffuse cloud conditions as a starting point. 

\begin{table} [! h]
\center
\caption{Assumed solar initial elemental abundances \citep{asplund}.}
\label{tab:initial}
\setlength{\tabcolsep}{10pt}
\begin{tabular} {c c} 
\hline \hline \\
Species  & Abundances \\
\hline \\ [-1ex ] 
H & 1.0 \\
He  &  $8.5\times10^{-2}$ \\
N   &  $6.8\times10^{-5}$ \\
O    & $4.9\times10^{-4}$ \\
$\mathrm{C^+}$  &  $2.7\times10^{-4}$ \\
$\mathrm{S^+}$ &  $1.3\times10^{-5}$ \\
$\mathrm{Si^+}$ &  $3.2\times10^{-5}$ \\
$\mathrm{Fe^+}$ & $3.2\times10^{-5}$ \\
$\mathrm{Na^+}$ & $1.7\times10^{-6}$ \\
$\mathrm{Mg^+}$ &  $3.9\times10^{-5}$ \\
$\mathrm{Cl^+}$ &  $3.2\times10^{-7}$ \\
$\mathrm{P^+}$ &  $2.6\times10^{-7}$ \\
$\mathrm{F^+}$ &  $3.6\times10^{-8}$ \\
\hline
\end{tabular} 
\end{table}

The phosphorus chemical network, that has been already used in previous studies \citep{fontani, rivilla16}, has been extended with new available information in the literature (new reactions, updated reaction rates, desorption energies etc.). In particular, chemical reactions of several P-bearing species, such as PN, PO, HCP, CP and  $\mathrm{PH_3}$ were included and/or updated in our chemical network. The reaction rates were taken from the online chemical databases KInetic Database for Astrochemistry 
\citep[KIDA]{wakelam}\footnote{http://kida.obs.u-bordeaux1.fr} and the UMIST Database for  Astrochemistry \citep[UDfA]{McElroy}\footnote{http://udfa.ajmarkwick.net/index.php?mode=species}, as well as from numerous previous papers \citep{thorne84, adams, millar91, anicich93, charnley94,  jimenez18}.
In particular we have included several reactions involving the formation and destruction of $\mathrm{PH_n}$ $(n=1,2,3)$ and their cationic species from \cite{charnley94} and \cite{anicich93}, along with the chemical network proposed by \cite{thorne84} that contains production and loss routes for P, PO, $\mathrm{P^+}$, $\mathrm{PO^+}$, $\mathrm{PH^+}$, $\mathrm{HPO^+}$ and $\mathrm{H_2PO^+}$. In addition, we extended the PN chemical network based on the work by \cite{millar87}, and we took into account the  gas-phase reaction $\mathrm{P + OH \rightarrow PO + H}$ proposed by \cite{jimenez18}, as well as two formation routes of PN in the gas-phase $\mathrm{N + CP \rightarrow PN +C}$ and $\mathrm{P+ CN \rightarrow PN +C}$ by \cite{agundez07}. Finally, we included the photodissociation reactions of PN, PO, HCP, and $\mathrm{PH_n}$ $(n=1,2,3)$ based on the reaction rates given in KIDA and UDfA. The reaction rates of the photodissociation of $\mathrm{PH_n}$ were assumed to be equal to the analogous reactions for $\mathrm{NH_n}$.

Concerning the chemistry taking place on grain surfaces, we have taken into account the hydrogenation reactions of P-bearing species (where the letter $``g"$ denotes a grain surface species) as well as their corresponding desorption reactions:
\begin{itemize}
\item{$\mathrm{gH + gP \rightarrow gPH}$}, \\
\item $\mathrm{gH + gPH \rightarrow gPH_2}$, \\
\item $\mathrm{gH + gPH_2 \rightarrow gPH_3}$. \\
\end{itemize}
The desorption energy of $\mathrm{PH_3}$ was calculated based on the one of $\mathrm{NH_3}$ and accounts to $\sim5800 \, \mathrm{K}$. This corresponds to an evaporation temperature of $\sim100 \, \mathrm{K}$, which is in good agreement with the value proposed by \cite{turner1990} ($\sim90 \, \mathrm{K}$)\footnote{The evaporation temperature describes the temperature at which a given species starts to desorb thermally.}.
The reactive desorption efficiency in our chemical model is set equal to 1\%. An increased reactive desorption of 10\% changes the predicted abundances of the aforementioned P-bearing molecules by less than a factor of 2. Another nonthermal desorption mechanism, that is included in our model, is the cosmic-ray desorption, which is fully described in \cite{hasegawa93}. Based on this study, dust grains are heated upon impact with cosmic-rays reaching a peak temperature $T_{\mathrm{dust}}$ of 70 K, which subsequently leads to preferential desorption of molecules from grain surfaces. This type of desorption is however negligible in diffuse clouds, where photodesorption dominates. In our model we adopt for all species a photodesorption rate of $3 \times 10^{-3}$ molecules per incident UV photon, as it was determined in \cite{oeberg2007} based on laboratory measurements of pure CO ice. This stands in good agreement with the photodesorption yield of $\sim10^{-3}$ molecules/UV photon found for other species, such as $\mathrm{H_2O}$, $\mathrm{O_2}$ and $\mathrm{CH_4}$ \citep{oeberg2009, fayolle2013, dupuy2017}.

\subsection{Comparison to observations}
In order to reproduce the observed abundances of HNC, CN, CS and CO in every cloud towards the line of sight to B0355+508 we produce a grid of models applying typical physical conditions for diffuse/translucent gas \citep{snow, thiel}. We note here that, since our chemical model is not treating isotopic species, we are using as a reference for our comparison the main species $\mathrm{^{12}CO}$ and $\mathrm{C^{32}S}$, instead of $\mathrm{^{13}CO}$ and $\mathrm{C^{34}S}$. For the fractional abundances of $\mathrm{^{12}CO}$ and $\mathrm{C^{32}S}$, we are adopting the column densities determined in \cite{liszt98} and \cite{liszt02}. In addition, for the clouds at $-14 \,\mathrm{km \, s^{-1}}$ and $-4 \,\mathrm{km \, s^{-1}}$ we use for CN and CS the upper limits derived in this work ($N(\mathrm{CN}) < 10^{12} \, \mathrm{cm^{-2}}$) and in \cite{liszt02}. The parameter space that we investigate is listed below:
\begin{itemize}
\item $n\mathrm{(H)} = 100 - 1000\, \mathrm{cm^{-3}}$, spacing of $100\, \mathrm{cm^{-3}}$,  \\
\item $A_V = 1-5 \, \mathrm{mag}$,  spacing of $1 \, \mathrm{mag}$, \\
\item $T_{\mathrm{gas}}=20-100 \, \mathrm{K}$,  spacing of $10 \, \mathrm{K}$. \\
\end{itemize}
The chemical evolution in each model is simulated over $10^7$ yrs (100 timesteps) by assuming static physical conditions. For the cosmic-ray ionisation rate $\zeta(\mathrm{CR})$ we use a value of  $1.7 \times 10^{-16} \, \mathrm{s^{-1}}$, as it was derived in \cite{indriolo} (see Section \ref{P-stuff} for further explanation). This also corresponds to the values applied in \cite{godard14} and \cite{lepetit04}, where the best-fit models provided a cosmic-ray ionisation rate of $10^{-16} \, \mathrm{s^{-1}}$ and $2.5\times 10^{-16} \, \mathrm{s^{-1}}$, respectively. 

Given the above parameter space, we calculate the level of disagreement $D(t,r)$ between modeled and observed abundances (for the species HNC, CN, CS and CO), which, following \cite{wakelam_herbst} and \cite{vasyunin17}, we define as

\begin{eqnarray} \label{agreement}
D(t,r) = \sum_{j=1}^{N_{\mathrm{species}}} \lvert\log(x^{j}_{\mathrm{mod}}(t,r)) - \log(x^{j}_{\mathrm{obs}})\rvert, 
\end{eqnarray}

\noindent{with $r=(n\mathrm{(H)}, A_V, T_{\mathrm{gas}})$ and $x^{j}_{\mathrm{obs, \, mod}}$ being the observed and modeled abundance of species $j$, respectively.} We then determine the minimal value of $D(t,r)$, (noted as $D_{\mathrm{min}}(t,r)$), that corresponds to the best-fit model; the parameters $(t,r)$ provided by the best-fit model are the ones giving the smallest deviation between observations and predictions.  The smaller the $D_{\mathrm{min}}(t,r)$, the better the agreement. 

According to \cite{pety} and references therein, the clouds with $\varv_{\mathrm{LSR}} = -10 \, \mathrm{km \, s^{-1}}$ and $-17 \, \mathrm{km \, s^{-1}}$ towards B0355+508, are showing strong  $\mathrm{^{12}CO}$ emission lines that originate from dense regions  (with $n\mathrm{(H)}=300-500 \, \mathrm{cm^{-3}}$ and  $N(\mathrm{H_2})>10^{21} \, \mathrm{cm^{-2}}$) which are right outside the synthesized beam (combination of the 30m and PdBI telescopes), but still within the IRAM 30m beam \citep{pety}.
Based on the CO (2-1) maps shown in \cite{pety} with a 22\arcsec and 5.8\arcsec resolution, the cloud at  $\varv_{\mathrm{LSR}} = -10 \,\mathrm{km \, s^{-1}}$ shows the most pronounced, dense sub-structure. This is confirmed by the fact that this particular cloud component produces the most detectable amounts of observed species, in the data presented in this work and previous works by \cite{liszt01} and \cite{liszt00}, thus being chemically the most complex one. The component at $\varv_{\mathrm{LSR}}  = -8 \,\mathrm{km \, s^{-1}}$ shows a similar structure as the one at $\varv_{\mathrm{LSR}}  = -10 \,\mathrm{km \, s^{-1}}$, incorporating a dense region as well. \cite{pety} suggests that the two components are part of the same cloud, even though they are distinguishable in absorption and show different levels of chemical complexity. With a higher spatial resolution of 5.8\arcsec, the CO emission at  $\varv_{\mathrm{LSR}}  = -8 \,\mathrm{km \, s^{-1}}$ is separated from the one at  $\varv_{\mathrm{LSR}}  = -10 \,\mathrm{km \, s^{-1}}$ and is clearly visible.
The diffuse gas seen at velocities $ -14 \,\mathrm{km \, s^{-1}}$ and $-4 \,\mathrm{km \, s^{-1}}$ shows barely any $\mathrm{^{13}CO}$ \cite[and this work]{liszt98}  and $\mathrm{^{12}CO}$ \citep{pety, liszt98} in emission, which suggests that the density of these clouds is too low to sufficiently excite CO. \cite{pety} estimated a low-moderate density of these clouds to be $\sim64-256 \, \mathrm{cm^{-3}}$ with $A_V < 2$ mag. 

Since we are performing single-dish observations with a beam of $\sim22\arcsec$, we also cover the high density regions that produce a significant $\mathrm{^{12}CO}$ emission \citep{pety}. For this reason  we constrain our grid of models to high densities of  $\geq 300 \, \mathrm{cm^{-3}}$  for the denser clouds ($-8,\, -10, \, -17 \, \mathrm{km \, s^{-1}}$), while for the low-density objects  at $-14 \, \mathrm{km \, s^{-1}}$ and $-4 \, \mathrm{km \, s^{-1}}$ we restrict our input parameters to $n\mathrm{(H)} \leq 200 \, \mathrm{cm^{-3}}$  and $A_V < 2$ mag. The calculation of the molecular abundances was done with respect to the $\mathrm{H_2}$ column densities ($N(\mathrm{H_2}) \sim4-5 \times 10^{20} \, \mathrm{cm^{-2}}$), that were derived towards every cloud by \cite{liszt18}. However, our model provides the fractional abundance of a species $X$ with respect to the total number of hydrogen nuclei, as in $n(\mathrm{X})/n(\mathrm{H})$, with the total volume density of hydrogen defined as $n\mathrm{(H)} = n\mathrm{(H \, I + 2 \cdot H_2)}$. The surface mobility parameters that we set as default values in our model (see Section \ref{surface_chemistry}) enable a fast and effective formation of $\mathrm{H_2}$ on the surfaces of grains. At the end of our simulations (at $t=10^7$ yrs), the $\mathrm{H_2}$ abundance reaches a value of $40\%-50\%$ (depending on the set of parameters). This means that almost the entire hydrogen is predicted to be in its molecular form at the late phases of the chemical evolution. Following this consideration, we divide all the observed abundances and their upper limits mentioned in this paper by a factor of 2, since $n\mathrm{(X)}/n\mathrm{(H \, I + 2 \cdot H_2)} \simeq n\mathrm{(X)}/2 \cdot n\mathrm{(H_2)} = 0.5 \cdot n\mathrm{(X)}/n\mathrm{(H_2)}$. We note that this expression applies to the high density parts of the clouds and does not account for the low density (and H I rich) gas along the line of sight.

\begin{figure*}[h]
	\centering
 	\includegraphics[width =1.0\textwidth]{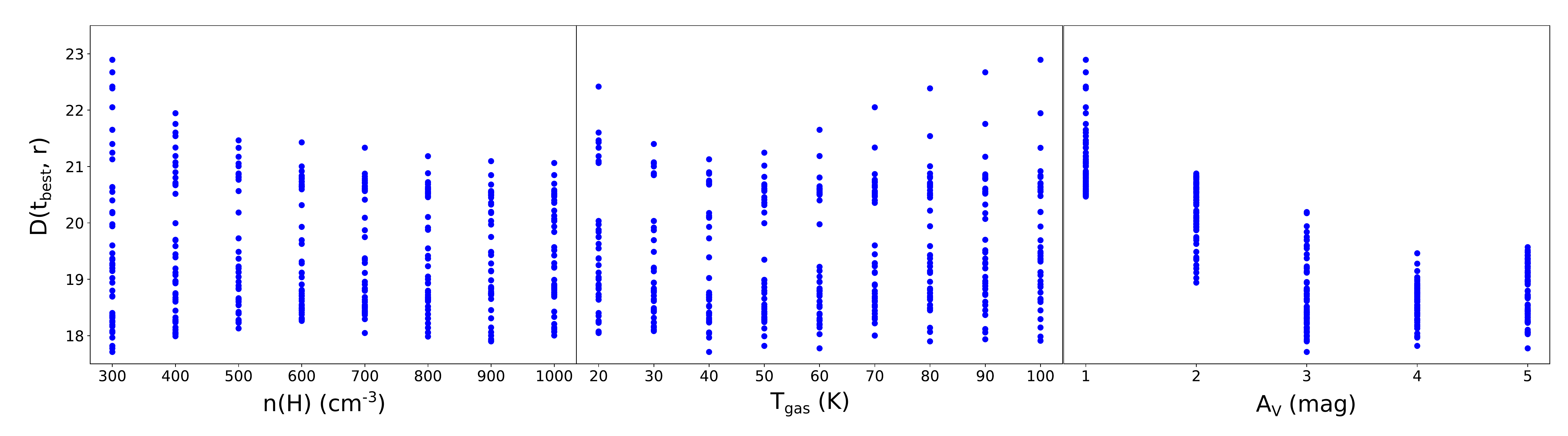}
	\caption{Results of the grid of models applying typical conditions for diffuse/translucent clouds in order to reproduce the observations towards the cloud at $\varv_{\mathrm{LSR}} = -17 \,\mathrm{km \, s^{-1}}$.  The deviation between observations and model at the time of best agreement $t_{\mathrm{best}}$ is given by  $D(t_{\mathrm{best}},r)$, which is plotted versus the density, temperature and visual extinction. 	
	The best-fit model is given at a time $t_{\mathrm{best}} = 6.2\times 10^6  \, \mathrm{yrs}$ and has the following parameters: $(n\mathrm{(H)}, A_V, T_{\mathrm{gas}}) = (300 \, \mathrm{cm^{-3}}, \, 3 \, \mathrm{mag}, \, 40 \, \mathrm{K})$.  Between an $A_V$ of 1 and 3 mag the minimal level of disagreement $D(t_{\mathrm{best}},r)$ drops down by 13\%.}
	\label{Fig:best model}
\end{figure*}

Figure \ref{Fig:best model} shows the results of the grid of models that was applied to reproduce the observations towards $\varv_{\mathrm{LSR}} = -17 \,\mathrm{km \, s^{-1}}$. In particular, we plot $D(t_{\mathrm{best}},r)$, which describes the deviation between observed and modeled abundances at the time of best agreement $t_{\mathrm{best}}$, versus the density, temperature and visual extinction. 
Between an $A_V$ of 1 and 3 mag the smallest level of disagreement $D(t_{\mathrm{best}},r)$ reduces by 13\%. The main discrepancy between observed and modeled abundances at low $A_V$ comes from the fact that a high visual extinction results in higher molecular abundances and is therefore able to reproduce the chemical complexity seen towards the translucent clouds. For models with $A_V > 3 \, \mathrm{mag}$ the minimal $D(t_{\mathrm{best}},r)$ is barely changing (less than 1\% of increase). The smallest $D(t_{\mathrm{best}},r)$ increases with respect to the density and temperature up to 3\% and 2\%, respectively. This is a clear indication that the most influential physical parameter in our analysis is the visual extinction. 

\begin{figure*}[h]
	\centering
 	\includegraphics[width = 0.7\textwidth]{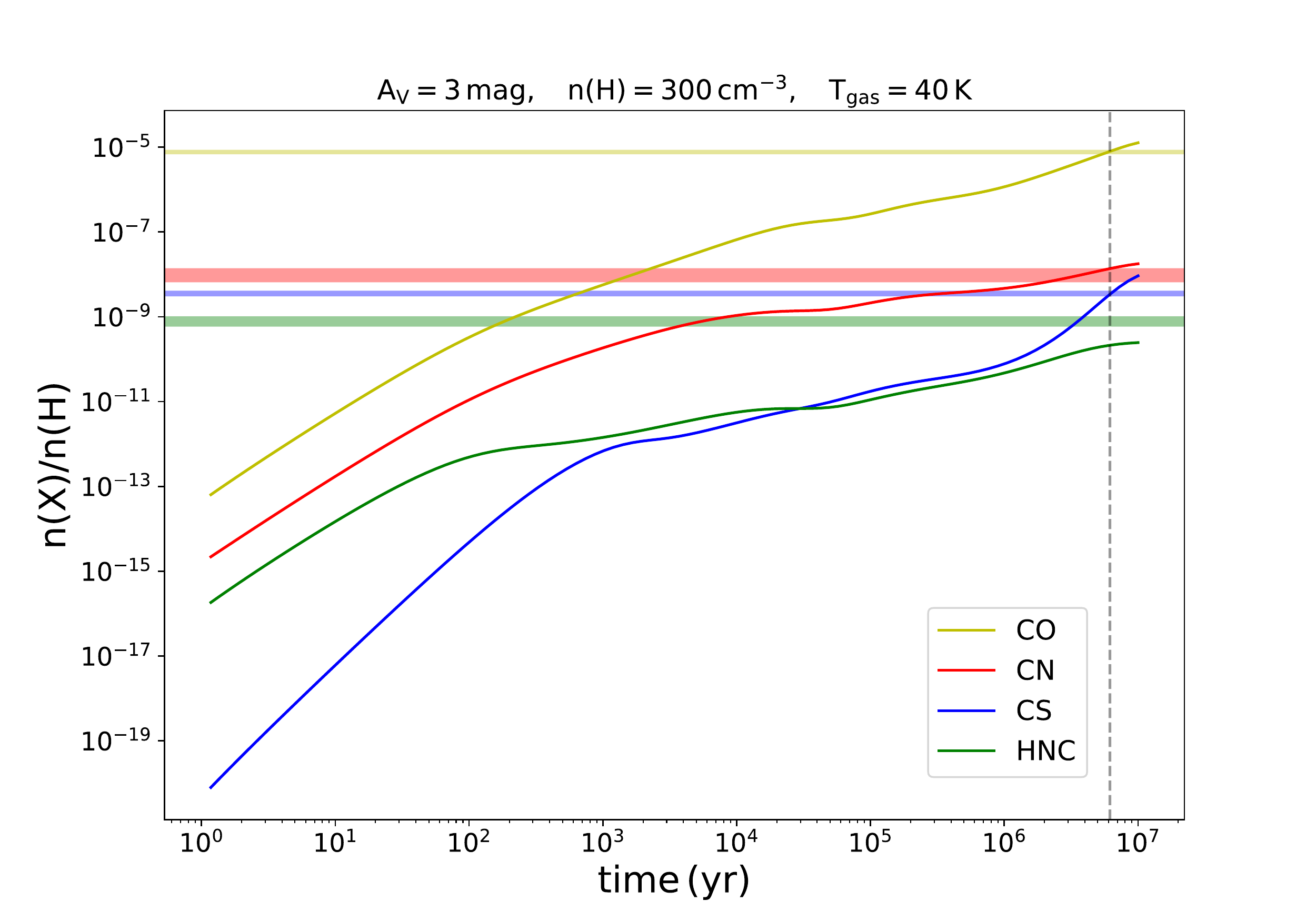}
	\caption{Chemical evolution of the abundances of CO, CN, CS and HNC over $10^7$ yrs predicted by our best-fit  model with the parameters $(n\mathrm{(H)}, A_V, T_{\mathrm{gas}}) = (300 \, \mathrm{cm^{-3}}, \, 3 \, \mathrm{mag}, \, 40 \, \mathrm{K})$. The coloured horizontal bands correspond to the observed abundances towards the cloud with $\varv_{\mathrm{LSR}}  = -17 \,\mathrm{km \, s^{-1}}$, including the inferred uncertainties. The vertical dashed line indicates the time of best agreement ($t=6.2 \times 10^6$ yrs) between observations and model results.}
	\label{Fig:best model2}
\end{figure*}

For the cloud component at $\varv_{\mathrm{LSR}} = -17 \,\mathrm{km \, s^{-1}}$ the best-fit model with $D_{\mathrm{min}}(t,r)$  is reached at a time $t_{\mathrm{best}} = 6.2\times 10^6  \, \mathrm{yrs}$  and has the parameters:  $r_{\mathrm{best}} = (n\mathrm{(H)}, A_V, T_{\mathrm{gas}}) = (300 \, \mathrm{cm^{-3}}, \, 3 \, \mathrm{mag}, \, 40 \, \mathrm{K})$. At $t_{\mathrm{best}} = 6.2\times 10^6  \, \mathrm{yrs}$ we also fulfill the assumption of  having most of hydrogen in molecular form, as the $\mathrm{H_2}$ abundance reaches a value of 0.45. Based on this model, we show in Figure \ref{Fig:best model2} the time dependent abundances of CO, CN, CS and HNC over $10^7$ yrs as well as the corresponding observed abundances towards $\varv_{\mathrm{LSR}}  = -17 \,\mathrm{km \, s^{-1}}$.  Our chemical model is reproducing the observed species CO, CN and CS very well within a factor of $\sim 1-1.4$ at $t_{\mathrm{best}}=6.2 \times 10^6$ yrs.  As it can be seen in Figure \ref{Fig:best model2}, the predicted abundances follow the order of the observed quantities. The most significant discrepancy is found in case of HNC, where the chemical model underestimates by a factor of 4 the observed abundance at the time of best agreement (see Table \ref{tab:comparison_model_obs}). According to the model, one of the main destruction mechanisms of HNC is: $\mathrm{C^{+} + HNC  \rightarrow C_2N^{+} + H}$. Based on the online chemical databases, its reaction rate remains uncertain. In UDfA the given reaction rate was determined theoretically \citep{leung}, and yields therefore a high uncertainty. Experimental studies of the above chemical route are still needed to make reliable predictions of the HNC abundance. 

\begin{table*} [! h]
\center
\caption{Observed abundances for the cloud component with $\varv_{\mathrm{LSR}}  = -17 \, \mathrm{km \, s^{-1}}$ and predictions of the species HNC, CO, CS and CN based on our best-fit model at the time of best agreement  $t= 6.2\times 10^6  \, \mathrm{yrs}$.  The last column lists the ratio of observed to predicted abundances.}
\label{tab:comparison_model_obs}
\setlength{\tabcolsep}{10pt}
\begin{tabular} {c c c c c } 
\hline \hline \\
Species & Observed   & Predicted   & Ratio\\
 &  Abundance  &  Abundance  &  (Observed/Predicted)\\
\hline \\ [-1ex ] 
HNC & $8.0(1.9)\times10^{-10}$ & $2.1\times10^{-10}$ &  3.8 \\

CN & $1.0(0.3)\times10^{-8}$ & $1.4\times10^{-8}$ & 0.7 \\

CS & $3.6(0.4)\times10^{-9}$ & $3.4\times10^{-9}$ &  1.1 \\

CO & $7.7(0.6)\times10^{-6}$ & $8.0\times10^{-6}$  & 1.0 \\
\hline
\end{tabular} 
\tablefoot{For the calculation of the observed abundances we used an $N(\mathrm{H_2})$ value of $4.30 \times 10^{20} \, \mathrm{cm^{-2}}$, as determined in \cite{liszt18}.}
\end{table*}

The same set of physical parameters provided also for the cloud with $\varv_{\mathrm{LSR}}  = -10 \,\mathrm{km \, s^{-1}}$ the smallest deviation with the observations. However, in this case the best-fit model gives a $D_{\mathrm{min}}(t,r)$, that is slightly larger by a factor of $\sim1.4$. The molecular abundances observed towards  $\varv_{\mathrm{LSR}}  = -8 \,\mathrm{km \, s^{-1}}$ are best reproduced with an $A_V$ of 5 mag, a density of  $400 \, \mathrm{cm^{-3}}$ and a  gas temperature of 40 K. The smallest level of disagreement between an $A_V$ of 3 and 5 mag differs by less than $1\%$.

For the two remaining clouds with  $\varv_{\mathrm{LSR}}  = -4 \,\mathrm{km \, s^{-1}}$ and $\varv_{\mathrm{LSR}} = -14 \,\mathrm{km \, s^{-1}}$ the best-fit model in both cases is given by  the parameters: $(n\mathrm{(H)}, A_V, T_{\mathrm{gas}}) = (200 \, \mathrm{cm^{-3}}, \, 1 \, \mathrm{mag}, \, 30 \, \mathrm{K})$ at a time $t_{\mathrm{best}} = 10^7  \, \mathrm{yrs}$. Here, the discrepancy in both clouds arises mostly from the fact that the model underestimates the CS abundance by a factor of $\sim 6-9$.  In our model, CS is being effectively destroyed via photodissociation due to the low $A_V$. 
Table \ref{tab:best-fit models} lists the best-fit parameters, that were determined towards every cloud component. 

We note that towards the same line of sight there have been  detections of several other molecules, as reported in \cite{liszt2008}. The best-fit model determined towards $\varv_{\mathrm{LSR}}  = -17 \,\mathrm{km \, s^{-1}}$ and $\varv_{\mathrm{LSR}} = -10 \,\mathrm{km \, s^{-1}}$ is able to reproduce within one order of magnitude the species OH, $\mathrm{C_2H}$, $\mathrm{H_2CO}$, $\mathrm{NH_3}$ and CH, while other species such as HCN, SO, $\mathrm{H_2S}$ and $\mathrm{C_3H_2}$ are strongly underestimated up to two orders of magnitude. This is a clear indication that the chemical network of certain molecules (other than P-bearing ones) still needs to be extended and updated. This however will be adressed in future work, as the present paper is focusing mainly on the P-bearing chemistry.

\begin{table*} [h!]
\centering
\caption{Set of physical parameters that give the best agreement between model results and observations towards every cloud component.}
\label{tab:best-fit models}
\setlength{\tabcolsep}{10pt}
\begin{tabular} {c c c c c} 
\hline \hline \\
Velocity  & $n(\mathrm{H})$ & $A_V$ & $T_{\mathrm{gas}}$ & $t_{\mathrm{best}}$ \\
($\mathrm{km \, s^{-1}}$) & ($\mathrm{cm^{-3}}$) & (mag) & (K) & ($10^6$ yrs) \\
\hline \\ [-1ex]
 -17 & 300 & 3 & 40 & 6.2 \\
 -14 & 200 & 1 &  30 & 10 \\
 -10 & 300 & 3 & 40 &  6.2 \\
 -8  & 400 & 5 & 40 & 2.3 \\
 -4 & 200 & 1 & 30 & 10 \\
\hline \\
\end{tabular} \\
\end{table*}

\section{Discussion: The chemistry of Phosphorus} \label{P-stuff}

Based on the above results we can conclude that the molecular abundances observed  at $\varv_{\mathrm{LSR}} = -8, \, -10, \, \, -17 \,\mathrm{km \, s^{-1}}$  can be best reproduced by a more "shielded" ($A_V>1 \, \mathrm{mag}$) interstellar medium that allows the build-up of molecules to occur more efficiently.  The resulting visual extinction $A_V$ of 3 mag should be viewed as an averaged value over the region covered by our beam ($\sim22\arcsec$). Within this region the denser clumps are most likely translucent in nature. Hence, the observed cloud components are probably heterogeneous clouds, incorporating diffuse and translucent material, filled with relatively abundant molecules. This result stands in good agreement with a study done by \cite{liszt17}, which involved modeling the CO formation and fractionation towards diffuse clouds. One of the main results was that strong $\mathrm{^{13}CO}$ absorption lines observed in the mm- and UV-range can be explained by higher densities ($\geq 256 \, \mathrm{cm^{-3}}$) and weaker radiation (and thus higher visual extinction), as already mentioned in Section \ref{results}. Our conclusions also agree well with the work done by \cite{thiel}, in which the physical and chemical structure of the gas along the line of sight to SgrB2(N) was studied; here, complex organic molecules, such as $\mathrm{NH_2CHO}$ and $\mathrm{CH_3CHO}$, were detected in the majority of the clouds, which at the same time proved to have relatively high visual extinctions ($A_V = 2.5 -5 \, \mathrm{mag}$ with $N(\mathrm{H_2})>10^{21} \, \mathrm{cm^{-2}}$), thus consisting mainly of translucent gas. According to \cite{thiel}  the column density of $\mathrm{H_2}$ that corresponds to an $A_V$ of 3 mag is $\sim 3\times 10^{21} \, \mathrm{cm^{-2}}$. This is also consistent with the study  by \cite{pety}, which states that the bright $\mathrm{^{12}CO}$ emission originates from dense regions with $N(\mathrm{H_2})>10^{21} \, \mathrm{cm^{-2}}$. 
The gas observed at velocities of $-14 \,\mathrm{km \, s^{-1}}$  and $-4 \,\mathrm{km \, s^{-1}}$ on the other hand, corresponds mainly to a "classical" diffuse cloud with a visual extinction of $\sim1$ mag according to the above analysis. These clouds also yielded the smallest amounts of the detected molecular abundances. 
Since chemical complexity seems to be favoured towards translucent rather than diffuse gas, we use for the following discussion as a reference model the one that provided the best fit towards the dense clouds with $\varv_{\mathrm{LSR}} =-17 \,\mathrm{km \, s^{-1}}$ and  $\varv_{\mathrm{LSR}} =-10 \,\mathrm{km \, s^{-1}}$.

According to our best-fit model,  $\mathrm{P^+}$ has a gas-phase abundance of $1.8\times10^{-7}$ at the end of our simulations, being a factor of $\sim1.4$ lower than its cosmic value, which indicates little depletion taking place. The main reservoir of phosphorus other than $\mathrm{P^+}$ is atomic P, having an abundance of $7.4\times10^{-8}$ at $10^7$ yrs.  Atomic P is formed mainly through the electronic recombination of $\mathrm{P^+}$. For our models with elevated densities ($10^3 \, \mathrm{cm^{-3}}$) we reach high elemental depletion (such as for $\mathrm{C^+}$, $\mathrm{S^+}$ and $\mathrm{P^+}$) after running the code for $10^7$ yrs.  This is consistent with the results presented by \cite{fuente19}, who show significant depletion of C, O and S happening already towards translucent material at the edge of molecular clouds (3-10 mag) with $1-5\times10^3 \, \mathrm{cm^{-3}}$. In the Appendix \ref{depletion} we investigate further the expected depletion of phosphorus when transitioning from diffuse- to dense-cloud conditions. We find that there is a significant depletion of atomic P on dust grains after the final density of $10^5 \, \mathrm{cm^{-3}}$ is reached. This in turn leads to a strong increase of $\mathrm{gPH_3}$, that becomes the main carrier of phosphorus in the dense phase. We also find a considerable decrease of the molecules HCP, CP, PN, PO and $\mathrm{PH_3}$ due to freeze-out on grains and their destruction route with $\mathrm{H_3^+}$ after the final density is attained at $t \sim 10^6-10^7$ yrs.

The most abundant P-bearing molecules in the gas-phase are HCP and CP, with maximal abundances of $3.4\times 10^{-10}$ and $2.1\times 10^{-10}$, respectively\footnote{The maximal abundances of the P-bearing species are reached at the end of our chemical simulations, at $t=10^7$ yrs. These abundances barely differ from the abundances at $t=6.2 \times 10^6$ yrs, which is the time of the best agreement with the observations.}. The formation and destruction pathways of both HCP and CP are strongly related to the electron fraction, as they are mainly produced (throughout the entire chemical evolution) by dissociative recombination of the protonated species $\mathrm{PCH_2^+}$ and destroyed by reacting with $\mathrm{C^{+}}$, the main carrier of positive charge in diffuse clouds. 
Two additional P-bearing species, that are predicted by our model to have "observable" abundances in the gas phase are PN and PO, with respective maximal abundances of $4.8\times 10^{-11}$ and $1.4\times10^{-11}$.  The most productive formation pathways for PN start with $\mathrm{P + CN \rightarrow PN + C}$, $\mathrm{N+PH \rightarrow PN + H}$ and end with $\mathrm{N + CP \rightarrow PN +C}$. In the late stage of evolution ($\sim0.5 \times10^6-10^7$ yrs) PN is primarily being destroyed by $\mathrm{He^{+} + PN \rightarrow P^{+} + N + He}$.  
The species PO is mainly produced over the entire chemical evolution of $10^7$ yrs by the dissociative recombination of $\mathrm{HPO^+}$:  $\mathrm{HPO^{+} + e^{-} \rightarrow PO + H}$, and is mostly destroyed by reactions with $\mathrm{C^+}$ and $\mathrm{H^+}$. On the other hand, $\mathrm{HPO^+}$ is efficiently formed via $\mathrm{P^{+} + H_{2}O \rightarrow HPO^{+} + H}$\footnote{The species $\mathrm{H_2O}$ is formed efficiently on dust grains ($\mathrm{gH + gOH \rightarrow H_2O}$) in the first $10^3$ yrs, while it is effectively produced via desorption $\mathrm{gH_2O \rightarrow H_2O}$ at late times. Our best-fit model produces a maximal $\mathrm{H_2O}$ abundance in the gas phase of $2.3\times10^{-8}$.}. 
An additional reaction that becomes relevant at progressive times ($\sim10^6-10^7$ yrs) is $\mathrm{O + PH \rightarrow PO + H}$ with a $\sim10\%$ reaction significance.

\begin{figure*}[h]
	\centering
 	\includegraphics[width = 0.7\textwidth]{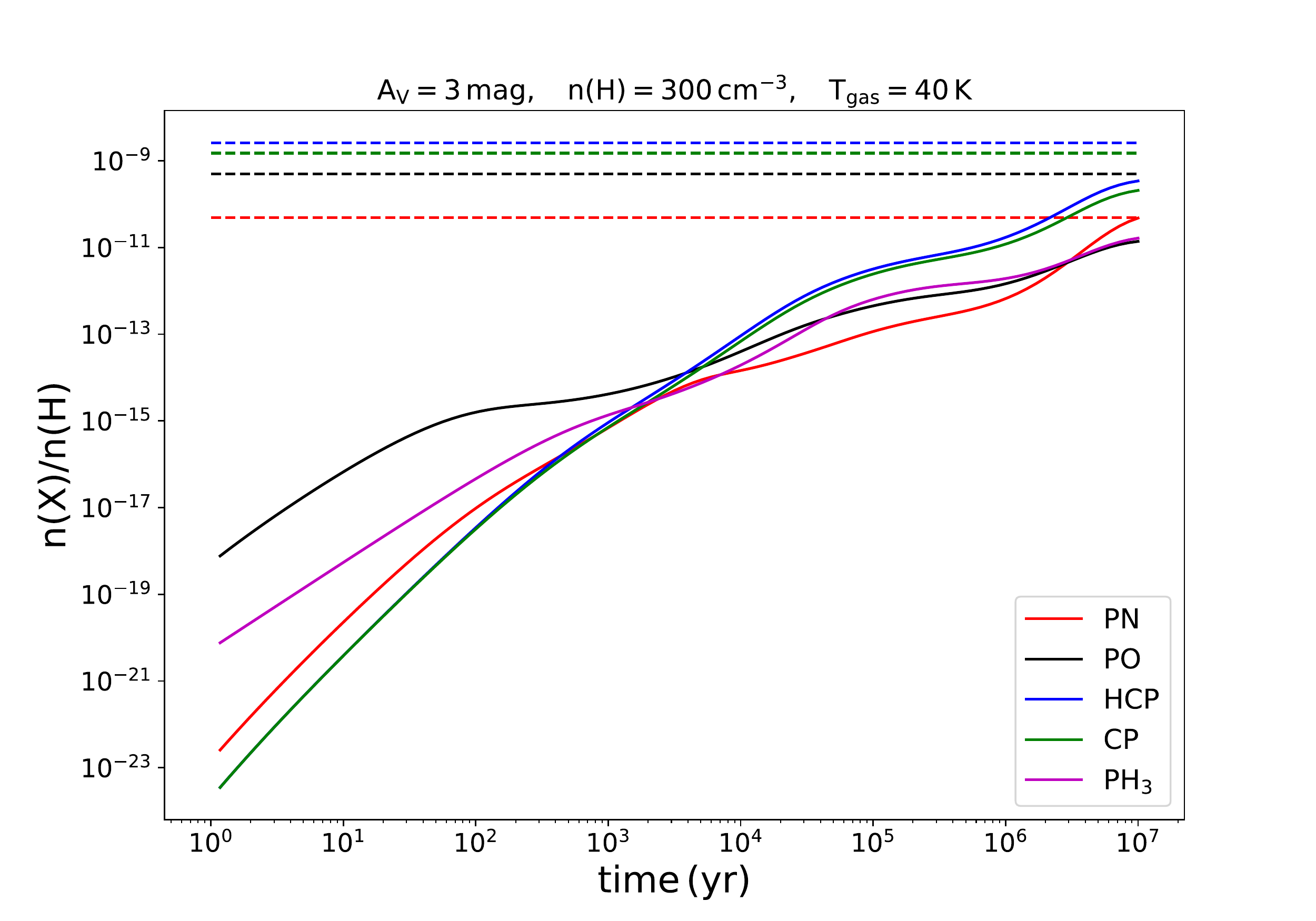}
	\caption{Variation of the predicted abundances of PN, PO, HCP, CP and $\mathrm{PH_3}$ over $10^7$ yrs in our best-fit model. The dashed lines represent the 3$\sigma$ upper limits derived from the observations at $\varv_{\mathrm{LSR}}  = -17 \, \mathrm{km \, s^{-1}}$. In case of PO we use as an upper limit the one at $5 \times 10^{-10}$ (see Table \ref{tab:comparison_prediction_upper_limits}).}
	\label{Fig:best model_Pstuff}
\end{figure*}

Another quite abundant P-bearing species in the gas-phase based on our best-fit model, is phosphine, $\mathrm{PH_3}$, with a maximal abundance of $\sim1.6 \times 10^{-11}$ at a late time of $10^7$ yrs.  We note here that the species PH is also predicted to be detectable with a maximal abundance of $\sim3.6 \times 10^{-11}$. Unlike PH however, $\mathrm{PH_3}$ has already been detected in circumstellar envelopes of evolved stars \citep{agundez14a}, indicating that it could be an important P-bearing species in interstellar environments such as diffuse and translucent clouds. Thus, we will focus in the following sections on the $\mathrm{PH_3}$  rather than the PH chemistry.
Based on our chemical model, $\mathrm{PH_3}$ is formed most efficiently on dust grains in the early phase, being released  to the gas-phase via reactive desorption: $\mathrm{gH + gPH_2 \rightarrow PH_3}$. Its formation proceeds after $1.4 \times 10^3$ yrs with the photodesorption process $\mathrm{gPH_3  \rightarrow PH_3}$ as the most effective reaction.
Since the evaporation temperature of $\mathrm{PH_3}$ lies at $\sim 100$ K,  the main mechanism driving the desorption of  $\mathrm{PH_3}$ at low temperatures is photodesorption (instead of thermal desorption). Switching off the photodesorption in our model, leads to a decrease of the $\mathrm{PH_3}$ gas-phase abundance of two orders of magnitude. Once in the gas-phase, $\mathrm{PH_3}$ is mostly destroyed by reactions with $\mathrm{C^+}$ and $\mathrm{H^+}$ as well as through the photodissociation reaction: $\mathrm{PH_3 + h\nu \rightarrow PH_2 + H}$. 
The most abundant species on grains is $\mathrm{gPH_3}$, with a maximal abundance of $7.2\times 10^{-10}$. Almost all the atomic P that depletes onto the dust grains reacts with gH and forms gPH ($\mathrm{gP + gH \rightarrow gPH}$), which subsequently forms $\mathrm{gPH_3}$  through further hydrogenation. Table \ref{tab:P_gasphase_chemistry} summarizes all the main formation and destruction pathways for the molecules PN, PO, HCP, CP and $\mathrm{PH_3}$ at three different times ($t = 10^3, 10^5, 10^7$ yrs).  The last column shows the significance of the given reaction in the total formation/destruction rate of the species of interest. 

\begin{table*} [h]
\centering
\caption{Main formation and destruction mechanisms for the species PN, PO, HCP, CP and $\mathrm{PH_3}$ based on the best-fit chemical model at times: $t=10^3, \, 10^5, \, 10^7$ yrs. The last column represents the share of the given reaction in the total formation/destruction rate of the corresponding species.}
\label{tab:P_gasphase_chemistry}
\setlength{\tabcolsep}{10pt}
\begin{tabular} {c c c c c} 
\hline \hline \\
Species & Time & Reaction type & Reaction  & Reaction importance   \\
 & (yrs) &  &  & (\%) \\
\hline \\ [-1ex]
PN &  $10^3$ & Formation &  $\mathrm{P + CN \rightarrow PN +C}$ & 47 \\
  &   & Formation &  $\mathrm{N + PH \rightarrow PN + H}$ & 25 \\  
   &   & Formation &  $\mathrm{N + PO \rightarrow PN + O}$ & 17 \\  [1ex]
    
   & $10^5$ & Formation & $\mathrm{N + CP \rightarrow PN + C}$ & 27 \\
    &    &  Destruction & $\mathrm{H^+ + PN \rightarrow PN^+ + H}$ & -27 \\
    &   & Destruction   &  $\mathrm{He^+ + PN \rightarrow P^+ + N + He}$ & -23 \\  
    &   &   Formation &  $\mathrm{N + PH \rightarrow PN + H}$ & 10 \\  [1ex]
    
    & $10^7$ & Destruction &  $\mathrm{He^+ + PN \rightarrow P^+ + N + He}$ & -41 \\
      &      & Formation & $\mathrm{N + CP \rightarrow PN +C}$  & 38 \\
        &    &  Destruction & $\mathrm{H^+ + PN \rightarrow PN^+ + H}$ & -8 \\ \hline \\
PO & $10^3$ & Formation &   $\mathrm{HPO^+ + e^- \rightarrow PO + H}$  & 48 \\      
      &   & Destruction &   $\mathrm{C^+ + PO \rightarrow PO^+ + C}$  & -44 \\ [1ex]
      
      & $10^5$ & Formation & $\mathrm{HPO^+ + e^- \rightarrow PO + H}$  & 47 \\
      &   & Destruction &  $\mathrm{H^+ + PO \rightarrow PO ^+ + H}$  & -37 \\
   &   & Destruction &  $\mathrm{C^+ + PO \rightarrow PO ^+ + C}$  & -9 \\  [1ex]
   
   &  $10^7$  & Formation  & $\mathrm{HPO^+ + e^- \rightarrow PO + H}$  & 36 \\
      &   & Destruction &  $\mathrm{C^+ + PO \rightarrow PO ^+ + C}$  & -26 \\
      &   & Formation  &  $\mathrm{O + PH \rightarrow PO + H}$  & 13 \\ \hline \\
HCP & $10^3$ & Formation &  $\mathrm{PCH_2^+  + e^- \rightarrow HCP + H}$ &  55 \\
 &  & Destruction &  $\mathrm{C^+  + HCP  \rightarrow CCP^+ + H}$ & -22 \\
   &  &  Destruction &  $\mathrm{C^+  + HCP  \rightarrow HCP^+ + C}$ & -22 \\  [1ex]

      & $10^5$ & Formation &  $\mathrm{PCH_2^+  + e^- \rightarrow HCP + H}$ &  50 \\
      &   & Destruction &  $\mathrm{H^+  + HCP  \rightarrow HCP^+ + H}$ &  -39 \\  [1ex]
      
    & $10^7$ & Formation & $\mathrm{PCH_2^+  + e^- \rightarrow HCP + H}$ &  50 \\
    &  & Destruction &  $\mathrm{C^+  + HCP  \rightarrow CCP^+ + H}$ & -17 \\
   &  &  Destruction &  $\mathrm{C^+  + HCP  \rightarrow HCP^+ + C}$ & -17 \\  \hline \\
   
 CP & $10^3$ & Destruction &  $\mathrm{C^+  + CP  \rightarrow CP^+ + C}$  &  -45 \\ 
      &  & Formation &  $\mathrm{PCH_2^+  + e^- \rightarrow CP + H_2}$ &  34 \\ [1ex]
 
     & $10^5$ & Destruction &  $\mathrm{H^+  + CP  \rightarrow CP^+ + H}$   &  -38 \\
     &    & Formation &  $\mathrm{PCH_2^+  + e^- \rightarrow CP + H_2}$  &  29 \\
     &   &  Formation & $\mathrm{HCP^+  + e^-  \rightarrow CP + H}$ & 18  \\ [1ex]
     
     & $10^7$ & Formation & $\mathrm{PCH_2^+  + e^- \rightarrow CP + H_2}$ &  35 \\
     &   & Destruction &  $\mathrm{C^+  + CP  \rightarrow CP^+ + C}$ &  -32 \\
     &   & Destruction &  $\mathrm{H^+  + CP  \rightarrow CP^+ + H}$  & -9 \\ \hline \\

$\mathrm{PH_3}$ &  $10^3$ & Destruction  & $\mathrm{C^+  + PH_3  \rightarrow PH_3^+ + C}$ & -45 \\
 & & Formation & $\mathrm{gH + gPH_2  \rightarrow PH_3}$ & 31 \\
 &  & Formation & $\mathrm{gPH_3  \rightarrow PH_3}$  & 23 \\ [1ex]

  &  $10^5$ & Formation & $\mathrm{gPH_3  \rightarrow PH_3}$  & 49 \\
  &   &  Destruction &  $\mathrm{H^+  + PH_3  \rightarrow PH_3^+ + H}$ &  -26 \\
  &   & Destruction  &  $\mathrm{C^+  + PH_3  \rightarrow PH_3^+ + C}$ &  -20 \\ [1ex]
  
  & $10^7$ & Formation &  $\mathrm{gPH_3  \rightarrow PH_3}$ & 49 \\
   &   & Destruction & $\mathrm{C^+  + PH_3  \rightarrow PH_3^+ + C}$ &  -36 \\
  &   & Destruction  & $\mathrm{PH_3  + h\nu \rightarrow PH_2 + H}$ &  -6 \\
\hline \\
\end{tabular} \\
\end{table*}

\begin{table*} [h]
\centering
\caption{Observed and predicted abundances at  time  $t= 10^7  \, \mathrm{yrs}$ for the species PO, PN, HCP, CP and $\mathrm{PH_3}$ given by our best-fit model.}
\label{tab:comparison_prediction_upper_limits}
\setlength{\tabcolsep}{10pt}
\begin{tabular} {c  c c} 
\hline \hline \\
Species  & Observed  & Predicted   \\
  &  Abundance & Abundance \\
\hline \\ [-1ex]
PN &    $<4.9 \times 10^{-11}$ &  $4.8 \times 10^{-11}$  \\
PO &    $<5.0 \times 10^{-10}$ &  $1.4 \times 10^{-11}$  \\
HCP &  $<2.6 \times 10^{-9}$ & $3.4 \times 10^{-10}$   \\
CP &  $<1.5 \times 10^{-9}$  & $2.1 \times 10^{-10}$  \\
$\mathrm{PH_3}$ &  - & $1.6 \times 10^{-11}$  \\
\hline \\
\end{tabular} \\
\tablefoot{The upper limits are 3$\mathrm{\sigma}$. For the calculation of the upper-limit-abundances we used an $N(\mathrm{H_2})$ value of $4.30 \times 10^{20} \, \mathrm{cm^{-2}}$ \citep{liszt18}. For PO we show the stringest upper limit. There are no observed data available for $\mathrm{PH_3}$.}\\
\end{table*}

Figure \ref{Fig:best model_Pstuff}  depicts the time dependent  abundances of PN, PO, HCP, CP and $\mathrm{PH_3}$ over $10^7$ yrs predicted by the best-fit model along with the computed 3$\sigma$ upper limits. The predicted abundance for PO lies a factor of $\sim40$ below the observational upper limit  at $t= 10^7  \, \mathrm{yrs}$, while
the current upper limits of HCP and CP are $\sim1$ order of magnitude higher than the model predictions. Finally, for PN the modeled abundance is almost reaching the observed value at the end of our simulations. This means, that in all cases the predicted abundances of P-bearing species are lower than the derived upper limits. Future observations of the ground energy transitions (1-0) will help us constrain even more these upper limits (see Section \ref{future} for further justification). 
Table \ref{tab:comparison_prediction_upper_limits} lists the predicted abundances of the above species given by our chemical model at $t= 10^7  \, \mathrm{yrs}$ along with the corresponding upper limits.

In the following discussion we focus on how deviations from our best-fit model can affect the P-bearing chemistry. In particular, we examine the dependence of the abundances of HCP, CP, PN, PO and $\mathrm{PH_3}$ on increasing visual extinction $A_V$, increasing cosmic-ray ionisation rate $\zeta(\mathrm{CR})$ as well as alternating the surface mobility constants (diffuse/desorption ratio $E_b/E_D$ and possibility of quantum tunneling for light species).

\subsection{Effects of visual extinction on the P-bearing chemistry}
\begin{figure*}[h]
	\centering
 	\includegraphics[width = 1\textwidth]{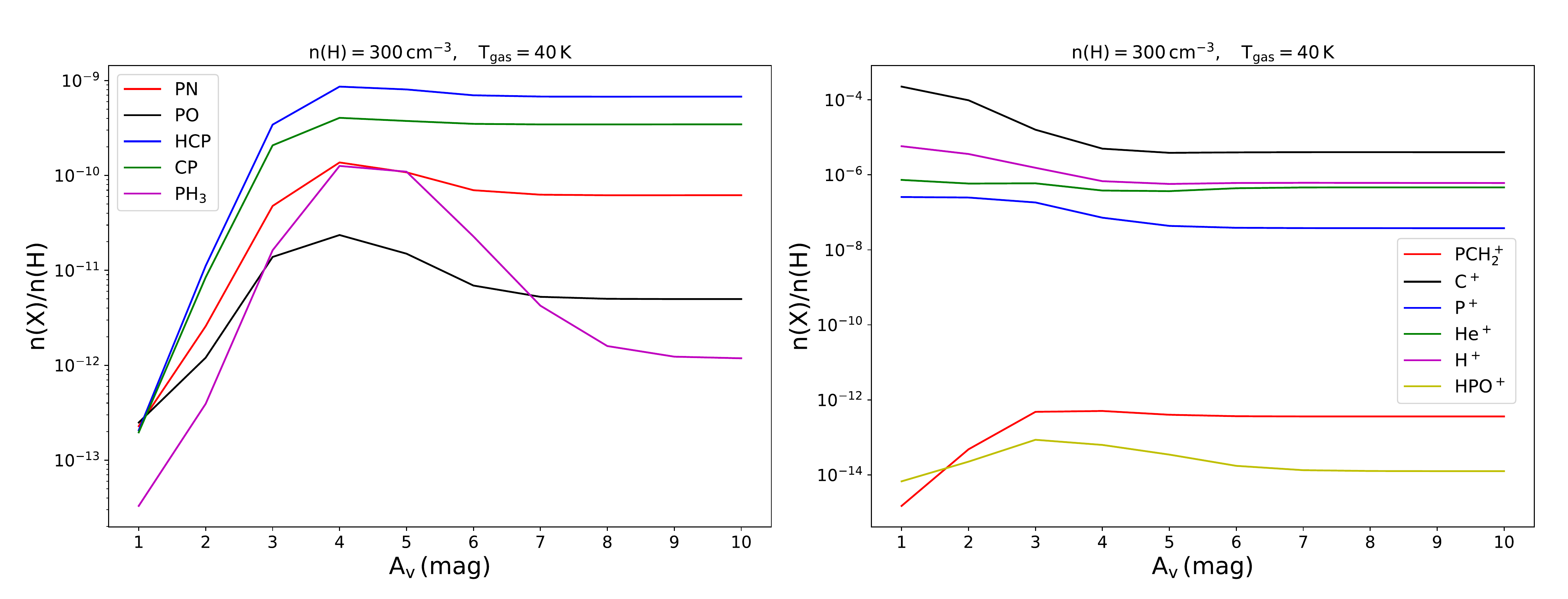}
	\caption{Predicted abundances of P-bearing molecules as a function of visual extinction $A_V$. The molecular abundances shown here are computed at $t=10^7$ yrs. The right panel illustrates the predicted abundances of $\mathrm{PCH_2^+}$, $\mathrm{C^+}$, $\mathrm{P^+}$, $\mathrm{He^+}$, $\mathrm{H^+}$, $\mathrm{HPO^+}$ as they are contributing the most to the formation and destruction of HCP, CP, PN, PO and $\mathrm{PH_3}$ (left panel).}
	\label{Fig:visual extinction}
\end{figure*}

In this Section we analyze how an increase of $A_V$ is affecting the predicted abundaces of P-bearing species. For this purpose we consider the parameters of the best-fit model with $n(H) = 300 \, \mathrm{cm^{-3}}$ and $T_{\mathrm{gas}} = 40 \, \mathrm{K}$, while varying the $A_V$ from 1 to 10 mag. By keeping the density constant, we avoid high levels of elemental depletion. The increase in visual extinction can then be explained by a figurative increase of the source's size.  
Figure \ref{Fig:visual extinction} (left panel) shows the predicted abundances of P-bearing species at the end of our simulations ($t=10^7$ yrs) under the effect of varying the visual exinction. All species reach a maximal abundance at an $A_V$ of 4 mag. The abundances of HCP,  CP and PN barely change for $A_V>4$ mag, while for the rest of the molecules the abundances drop; especially in case of $\mathrm{PH_3}$ we denote a substantial decrease of almost two orders of magnitude. As already mentioned in Section \ref{P-stuff}, the most effective formation process of $\mathrm{PH_3}$ is the photodesorption $\mathrm{gPH_3 \rightarrow PH_3}$. Thus, a high visual extinction attenuates the incoming UV-field and therefore the desorption of $\mathrm{gPH_3}$. 

In order to better understand the $A_V$ dependence of the remaining molecular abundances, we have plotted in Figure \ref{Fig:visual extinction} the predicted abundances of the species that mainly form and destroy HCP, CP, PN and PO (see Table \ref{tab:P_gasphase_chemistry}) as a function of the visual extinction. 
In particular, we have simulated the abundances of $\mathrm{PCH_2^+}$, $\mathrm{C^+}$, $\mathrm{P^+}$, $\mathrm{He^+}$, $\mathrm{H^+}$ as well as $\mathrm{HPO^+}$.
In case of HCP (and also CP) its abundance increases up to an $A_V$ of 4 mag and subsequently stays constant above that value. This behaviour is correlated with the increase of the $\mathrm{PCH_2^+}$ abundance up to an $A_V$ of 3 mag as well as the decrease of $\mathrm{C^+}$ up to a visual extinction of 4 mag. 
The species PO seems to be more strongly affected by the increasing $A_V$. Its abundance will also increase for $A_V \leq 4 \, \mathrm{mag}$ which again stands in correlation with the decrease of the $\mathrm{C^+}$ abundance (the main "destroyer" of PO), followed by a drop in abundance up to 7 mag. This on the other hand results from the decrease of the $\mathrm{HPO^+}$ abundance (the main precursor of PO) in the same $A_V$ range. 
An increase in $A_V$ will decrease the $\mathrm{P^+}$ abundance (due to the decrease of the total ionisation rate), as it can be seen in Figure \ref{Fig:visual extinction}. In addition, an enhanced $A_V$ is slightly decreasing the $\mathrm{H_2O}$ abundance (by a factor of 2), since the most effective formation for $\mathrm{H_2O}$ at late times is the photodesorption $\mathrm{gH_2O \rightarrow H_2O}$ (see footnote 7). Therefore, for higher $A_V$, both $\mathrm{P^+}$ and $\mathrm{H_2O}$ decrease, so that  $\mathrm{HPO^{+}}$ and subsequently PO reduce in abundance as well. 

\subsection{Effects of the cosmic-ray ionisation rate on the P-bearing chemistry} 
\begin{figure*}[h]
	\centering
 	\includegraphics[width = 1\textwidth]{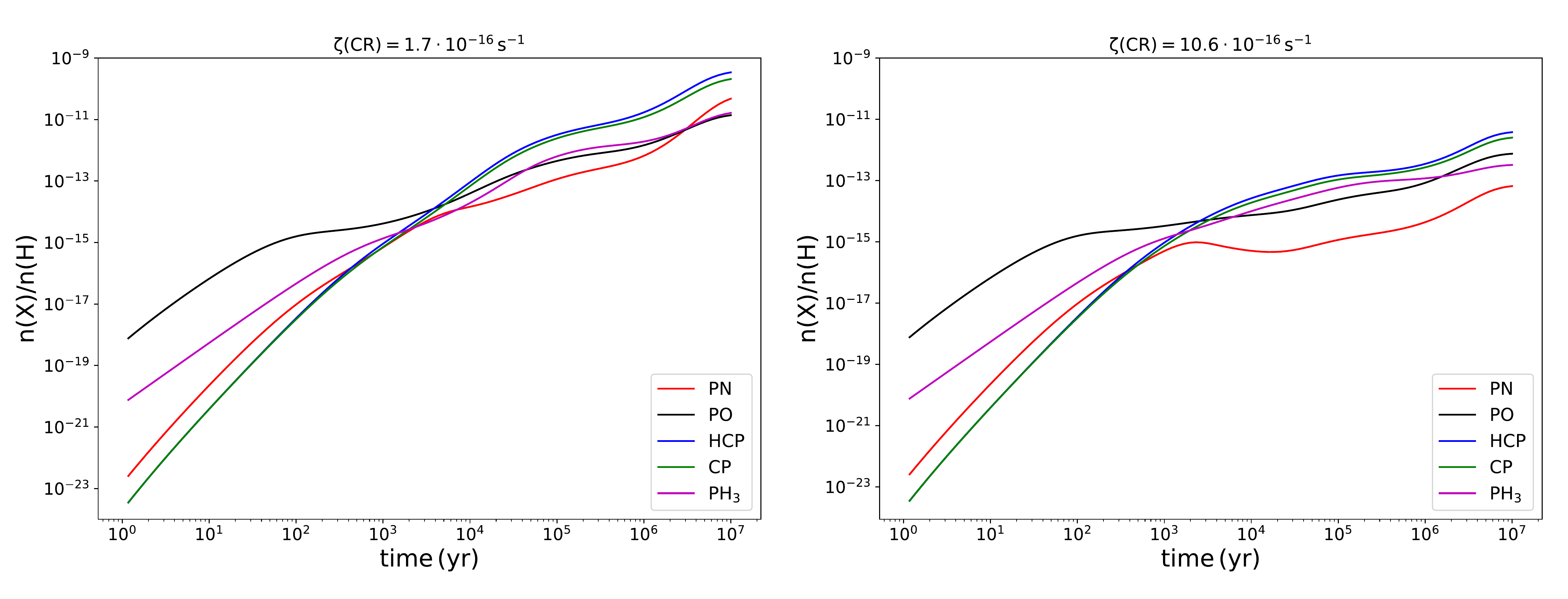}
	\caption{Chemical evolution of P-bearing molecules as a function of time under the effects of a cosmic-ray ionisation rate of $\zeta(\mathrm{CR}) = 1.7 \times 10^{-16} \, \mathrm{s^{-1}}$ (left panel) and $10.6 \times 10^{-16} \, \mathrm{s^{-1}}$ (right panel).}
	\label{Fig:zeta}
\end{figure*}

As already mentioned in Section \ref{model}, for all the applied models
we use for the cosmic-ray ionisation rate $\zeta(\mathrm{CR})$ a value of $1.7 \times 10^{-16} \, \mathrm{s^{-1}}$, as it was derived by \cite{indriolo}. This is also consistent with previous work in which diffuse and translucent clouds were studied as well \citep{fuente19, godard14, lepetit04}.
However, we should note here that in \cite{indriolo} several cosmic-ray ionisation rates were derived towards 50 diffuse line of sights, ranging from $1.7 \times 10^{-16} \, \mathrm{s^{-1}}$ to $10.6 \times 10^{-16} \, \mathrm{s^{-1}}$ with a mean value of  $3.5 \times 10^{-16} \, \mathrm{s^{-1}}$. 
Due to the complex and yet not fully known nature of our observed clouds we test our chemical model by also applying the elevated values of  $\zeta(\mathrm{CR}) = 3.5 \times 10^{-16} \, \mathrm{s^{-1}}$ and  $10.6 \times 10^{-16} \, \mathrm{s^{-1}}$ in order to examine the influence of the cosmic-ray ionisation rate on the P-bearing chemistry. 
As for the remaining parameters of the code (such as $A_V$ and $T_{\mathrm{gas}}$) we use the values given by our best-fit model (see Section \ref{model}). Figure \ref{Fig:zeta} shows the chemical evolution of the species PN, PO, HCP, CP and $\mathrm{PH_3}$ over $10^7$ yrs for a cosmic-ray ionisation rate of  $\zeta(\mathrm{CR}) = 1.7 \times 10^{-16} \, \mathrm{s^{-1}}$ and  $10.6 \times 10^{-16} \, \mathrm{s^{-1}}$ in the left and right panel, respectively.

\begin{table*} [! h]
\center
\caption{Predicted abundances of the species PN, PO, HCP, CP and $\mathrm{PH_3}$ at $t= 10^7$ yrs for three different cosmic-ray ionisation rates (see text for explanation).}
\label{tab:cosmic_rays}
\setlength{\tabcolsep}{10pt}
\begin{tabular} {c c c c} 
\hline \hline \\
Species & Predicted Abundances  &Predicted Abundances  & Predicted Abundances \\
 & ($\zeta(\mathrm{CR}) = 1.7 \times 10^{-16} \, \mathrm{s^{-1}} $) & ($\zeta(\mathrm{CR}) = 3.5 \times 10^{-16} \, \mathrm{s^{-1}} $) & ($\zeta(\mathrm{CR}) = 10.6 \times 10^{-16} \, \mathrm{s^{-1}} $) \\
\hline \\ [-1ex ] 
PN & $4.8\times10^{-11}$ & $2.9\times10^{-12}$  & $6.6\times10^{-14}$ \\

PO & $1.4\times10^{-11}$ & $4.4\times10^{-12}$ & $7.5\times10^{-13}$ \\

HCP & $3.4\times10^{-10}$ & $6.7\times10^{-11}$ & $3.8\times10^{-12}$\\

CP & $2.1\times10^{-10}$   & $4.2\times10^{-11}$ &$2.5\times10^{-12}$\\

$\mathrm{PH_3}$ & $1.6\times10^{-11}$   & $2.7\times10^{-12}$ &  $3.2\times10^{-13}$\\
\hline
\end{tabular} 
\end{table*}

Table \ref{tab:cosmic_rays} summarizes the predicted abundances of P-bearing species for the three different cosmic-ray ionisation rates given in \cite{indriolo}. As one can recognize, PN shows the most substantial decrease in abundance with increasing $\zeta(\mathrm{CR})$. From the lowest ($\zeta(\mathrm{CR}) = 1.7 \times 10^{-16} \, \mathrm{s^{-1}} $) to the highest ($\zeta(\mathrm{CR}) = 10.6 \times 10^{-16} \, \mathrm{s^{-1}} $) cosmic-ray ionisation rate, the PN abundance decreases by a factor of $\sim730$, while for HCP, CP and $\mathrm{PH_3}$ we have a drop by a factor of $\sim85$ and $\sim 50$, respectively.  As already mentioned, PN is heavily destroyed by $\mathrm{He^+}$ with a $\sim40 \%$ reaction significance. An increase of $\zeta(\mathrm{CR})$ up to a value of $10.6 \times 10^{-16} \, \mathrm{s^{-1}}$ is significantly enhancing the ionisation of He and H by a factor of $\sim 20$ and $\sim 30$, respectively via cosmic-ray-induced secondary UV photons: $\mathrm{He + CRP \rightarrow He^+ + e^-}$ and  $\mathrm{H + CRP \rightarrow H^+ + e^-}$ (the abundance of $\mathrm{C^+}$ increases by $\sim6$). Therefore, the destruction path with $\mathrm{H^+}$ becomes relevant for all P-bearing species showing a 10-40\% loss efficiency. The effect is the strongest in case of PN, because PN is mainly formed through CP which is drastically decreasing and is also efficiently destroyed by both $\mathrm{He^+}$ and $\mathrm{H^+}$.  The PO abundance is only reduced by a factor of $\sim20$ after increasing $\zeta(\mathrm{CR})$ up to $10.6 \times 10^{-16} \, \mathrm{s^{-1}} $, despite being heavily destroyed by $\mathrm{H^+}$. On the other hand, the significance of the dissociative recombination of $\mathrm{HPO^+}$ increases up to 50\% which in turn counterbalances the loss through $\mathrm{H^+}$. An increased $\zeta(\mathrm{CR})$ of $10.6 \times 10^{-16} \, \mathrm{s^{-1}}$ is enhancing the abundance of $\mathrm{P^+}$ up to $\sim2.5\times10^{-7}$, nearly reaching its cosmic value of $\sim2.6\times10^{-7}$ \citep{asplund}, while the abundance of atomic P decreases down to $\sim9.5\times10^{-9}$ via the enhanced reaction with $\mathrm{C^+}$ and  $\mathrm{H^+}$.

\subsection{Effects of the diffusion/desorption ratio on the P-bearing chemistry} \label{surface_chemistry}
The chemistry in the ISM is heavily influenced by the presence of dust grains \citep{caselli_cecca}. The mobility of the depleted species on the surface of dust grains depends on two mechanisms: thermal hopping and quantum tunneling for the lightest species H and $\mathrm{H_2}$ through potential barriers between surface sites \citep{hasegawa}. Without the possibility of tunneling, the species are not able to scan the grain surface quickly at low temperatures and the total mobility decreases. The parameters that strongly determine the surface chemistry are the diffusion/desorption energy ratio $E_b/E_D$ as well as the thickness of the potential barrier between adjacent sites. Based on previous studies \citep{hasegawa, ruffle, garrod}, \cite{vasyunin2013} proposed three different values for the $E_b/E_D$ ratio: 0.3, 0.5 and 0.77. In case of low ratios ($E_b/E_D=0.3$) we activate in our model the possibility of quantum tunneling for light species, while for the other two cases, surface mobility is only controlled by thermal hopping (and quantum tunneling is deactivated). The potential barriers are assumed to have rectangular shape and a thickness of 1 $\mathrm{\mathring{A}}$ \citep{vasyunin2013}. In our model we utilize the first set of parameters ($E_b/E_D=0.3$, with tunneling), nevertheless, since P-bearing chemistry is still highly uncertain, we examine how the remaining two sets of parameters ($E_b/E_D=0.5, \, 0.77$, no tunneling), influence the predicted abundances. Table \ref{tab:tunneling} lists the predictions of PN, PO, HCP, CP and $\mathrm{PH_3}$ as well as $\mathrm{H_2}$ at $t= 10^7$ yrs for the three different sets of surface mobility parameters proposed in \cite{vasyunin2013}.
As Table \ref{tab:tunneling} shows, the $\mathrm{H_2}$ abundance decreases by a factor of 4, by switching from set up 1 ($E_b/E_D=0.3$ with tunneling) to set up 2 ($E_b/E_D=0.5$ no tunneling), and finally experiences a dramatic drop of a factor 50 when increasing the $E_b/E_D$ up to 0.77 (overall change of a factor 200 between set up 1 and 3).

\begin{table*} [! h]
\center
\caption{Predicted abundances of the species PN, PO, HCP, CP and $\mathrm{PH_3}$ as well as $\mathrm{H_2}$  at  $t= 10^7$ yrs for three different sets of surface mobility parameters (see text for explanation).}
\label{tab:tunneling}
\setlength{\tabcolsep}{10pt}
\begin{tabular} {c c c c} 
\hline \hline \\
Species & Predicted Abundances  &Predicted Abundances  & Predicted Abundances \\
 & ($E_b/E_D=0.3$ with tunneling) & ($E_b/E_D=0.5$  no tunneling) & ($E_b/E_D=0.77$  no tunneling) \\
\hline \\ [-1ex ] 
PN & $4.8\times10^{-11}$ & $5.0\times10^{-13}$  & $1.6\times10^{-13}$ \\

PO & $1.4\times10^{-11}$ & $1.2\times10^{-12}$ &   $6.7\times10^{-13}$ \\

HCP & $3.4\times10^{-10}$ & $1.3\times10^{-11}$ & $4.6\times10^{-12}$\\

CP & $2.1\times10^{-10}$   & $9.5\times10^{-12}$ &  $3.2\times10^{-12}$\\

$\mathrm{PH_3}$ & $1.6\times10^{-11}$   & $1.8\times10^{-12}$ &  $1.2\times10^{-12}$\\

$\mathrm{H_2}$ & $4.8\times10^{-1}$   & $1.3\times10^{-1}$  &  $2.4\times10^{-3}$   \\
\hline
\end{tabular} 
\end{table*}

The reduction of the $\mathrm{H_2}$ abundance has a significant impact on the formation of $\mathrm{PCH_2^+}$ and PH, which affects the PN, PO, HCP and CP abundances through the following reactions:\\

\textbf{PN} 

\begin{itemize}
\item $\mathrm{P^+  + H_2 \rightarrow PH_2^+}$ \\
\item $\mathrm{PH_2^+  +  e^- \rightarrow PH + H}$ \\
\item $\mathrm{N + PH \rightarrow PN + H}$ \\
\end{itemize}

\textbf{PO} 

\begin{itemize}
\item $\mathrm{P^+  + H_2 \rightarrow PH_2^+}$ \\
\item $\mathrm{PH_2^+  +  e^- \rightarrow PH + H}$ \\
\item $\mathrm{O + PH \rightarrow PO + H}$ \\
\end{itemize}

\textbf{HCP} 

\begin{itemize}
\item $\mathrm{HCP^+  + H_2 \rightarrow PCH_2^+}$ \\
\item $\mathrm{PCH_2^+  +  e^- \rightarrow HCP + H}$ \\
\end{itemize}

\textbf{CP} 

\begin{itemize}
\item $\mathrm{HCP^+  + H_2 \rightarrow PCH_2^+}$ \\
\item $\mathrm{PCH_2^+  +  e^- \rightarrow CP + H_2}$ \\
\end{itemize}

\begin{figure*}[h]
	\centering
 	\includegraphics[width = 1\textwidth]{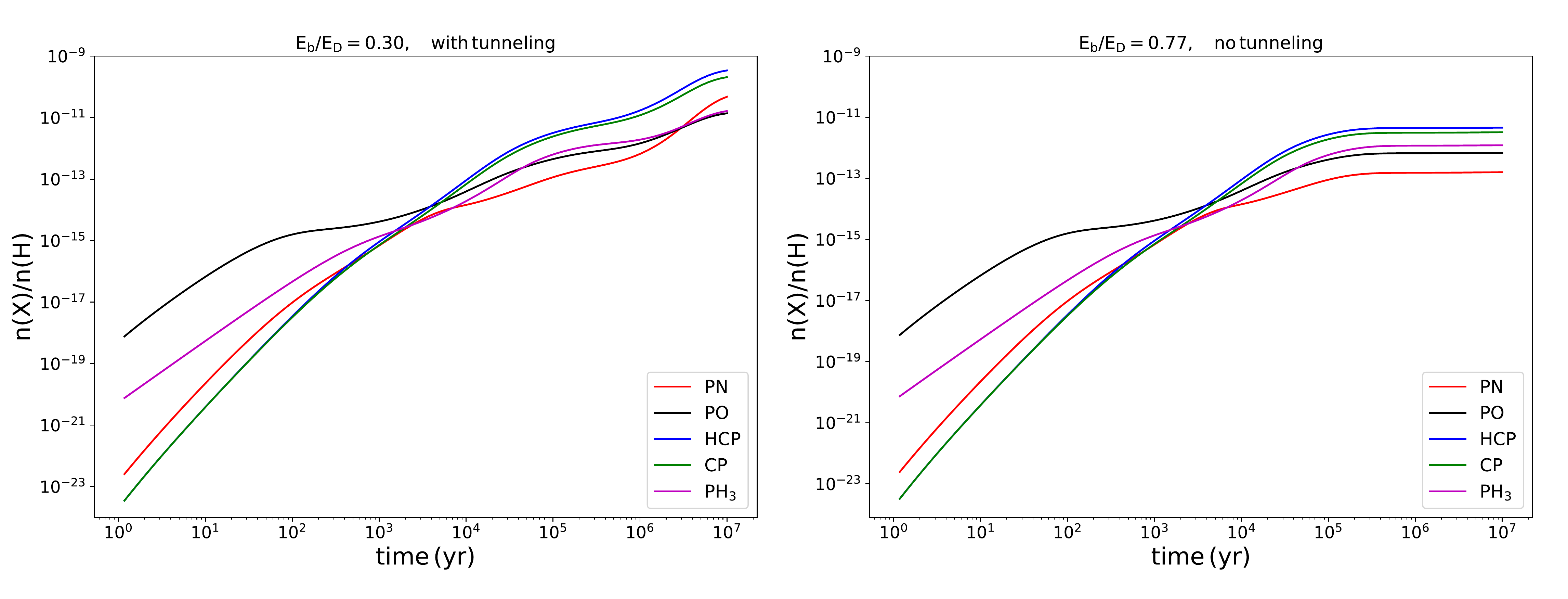}
	\caption{Chemical evolution of P-bearing molecules as a function of time for a diffusion/desorption ratio $E_b/E_D$ of 0.3 (with quantum tunneling) shown in the left panel and for a  $E_b/E_D$ of 0.77 (without quantum tunneling) in the right panel.}
	\label{Fig:diff_des}
\end{figure*}

Both PH and $\mathrm{PCH_2^+}$ decrease by a factor of $\sim 20$ when increasing the $E_b/E_D$ up to 0.77.
In addition, the abundance of $\mathrm{H^+}$ is increased by a factor of $\sim25$, since the reduction of $\mathrm{H_2}$ formation leads to more atomic hydrogen and subsequently also $\mathrm{H^+}$. The enhanced $\mathrm{H^+}$ abundance results in a stronger destruction of all P-bearing species through their reaction with $\mathrm{H^+}$. The species HCP and CP are also strongly affected by changing the surface mobility parameters, with an overall decrease of a factor $\sim70$ and $\sim65$ in abundance, respectively. In both cases the dissociative recombination of $\mathrm{PCH_2^+}$ is essential during the whole chemical evolution for the formation of HCP and CP showing a reaction significance of 30 to 99\%.  A decrease of $\mathrm{PCH_2^+}$ due to lower $\mathrm{H_2}$ abundance is therefore resulting in a reduced HCP and CP formation.  The largest effect is seen for PN, where a diffusion/desorption ratio of 0.77 and no quantum tunneling of light species reduces the PN abundance by a factor of 300 (see Figure \ref{Fig:diff_des}). 

Besides the effective loss through $\mathrm{H^+}$, the substantial decrease in PN is also related to the reduction of CP, which is the main precursor of PN at late times. In addition, the reaction $\mathrm{N + PH \rightarrow PN +H}$ is significant to the PN formation over the entire chemical evolution of $10^7$ yrs with a 10-50\% formation efficiency (for $E_b/E_D=0.77$ and no tunneling). This means that the reduction of the $\mathrm{H_2}$ abundance is decreasing PH, which in turn produces less PN. In case of PO however,  the change in abundance between the two extreme cases is just a factor of $\sim 20$. Here, the route $\mathrm{O +PH \rightarrow PO + H}$  increases in significance only up to 3\% at late times ($4\times 10^6- 10^7$ yrs), indicating that the decrease of PH will not considerably affect the PO production. Furthermore, the reduction of PO due to $\mathrm{H^+}$ is  compensated through its effective formation via the dissociative recombination of $\mathrm{HPO^+}$. Finally, the abundance of $\mathrm{PH_3}$ decreases only by a factor 13 in total when changing the surface chemistry constants.  Despite being heavily destroyed by $\mathrm{H^+}$, $\mathrm{PH_3}$ is still sufficiently formed through the photodesorption of $\mathrm{gPH_3}$.

\section{Future observations} \label{future}
Thanks to the sensitive observations (rms of $\sim 6$ mK)  of the (2-1) transitions of HCP, CP, PN and PO we were able to obtain good upper limits for the column densities and abundances of the above species (see Tables \ref{tab:upper_limits_PN_PO} and \ref{tab:comparison_prediction_upper_limits}) and thus constrain the P-chemistry. 
The observations of HNC, CN, CS and CO helped us put important constraints on the main physical parameters of the targeted diffuse/translucent clouds, i.e. the visual extinction, the density and the gas temperature. 
For the prospect of future observations we want to estimate the expected line instensities of the (1-0) transitions of HCP, CP, PN and PO (at $\sim40-65$ GHz) based on our new and improved diffuse-cloud model. Since the densities present in diffuse/translucent clouds are too low to show any collisional excitation ($T_{\mathrm{ex}} = T_{\mathrm{bg}} = 2.7 \, \mathrm{K}$), the (1-0) transitions are expected to be more strongly populated than the higher energy transition levels. For these calculations, we take into account that the emission of the blazar is non-thermal, meaning that the flux increases with decreasing frequency. In particular, we apply a  power law for the blazar's emission with $\frac{F}{F_0} = (\frac{\nu}{\nu_0})^{-\alpha}$, where $F$ is the flux, $\nu$ is the corresponding frequency and $\alpha$ is the spectral index.  By using the fluxes determined in \cite{agudo} at 3 and 1.3 mm we infer a spectral index of $\alpha \sim1.06$. Following this, we determine the flux at 7 mm to be $\sim11$ Jy, which in turn corresponds to a temperature of $\sim26$ K with a beam size of $17\arcsec$ (at 7 mm with the Green Bank Telescope).  As Table \ref{tab:expected_intens} shows, the derived peak intensities of the species PN, PO, HCP and CP  vary from 10 to 200 mK, making these lines "detectable" with radio telescopes, such as the Green Bank Telescope (GBT) and the Effelsberg Telescope. The capabilities of these instruments will allow us to reach rms levels down to 4 mK and enable possible detections up to a 50$\sigma$ level. The only exception is $\mathrm{PH_3}$ with a (1-0) transition at 266.944 GHz. The  flux of the background source at that frequency based on the above power law is equal to 1.91 Jy. This corresponds to a background temperature $T_{\mathrm{c}}$ of 0.4 K with a beam size of $9\arcsec$ (with the IRAM telescope), which in the end results in a very weak, non-detectable absorption line.

\begin{table*} [! h]
\center
\caption{Estimated absorption line intensities for the (1-0) transitions of HCP, CP,  PN and PO towards B0355+508 for $T_{\mathrm{ex}}=2.73 \,\mathrm{K}$, a FWHM linewidth of $\Delta \varv = 0.5 \, \mathrm{km \, s^{-1}}$ and based on the predicted abundances given by our best-fit model at $t=10^7$ yrs.}
\label{tab:expected_intens}
\setlength{\tabcolsep}{10pt}
\begin{tabular} {c c c c c c c c} 
\hline \hline \\ [-2ex]
Species & Transitions & $E_{\mathrm{up}}$  & Frequency & $A_\mathrm{ul}$ & $g_u$  & Estimated Intensities & References \\
& & (K) & (GHz) & ($\mathrm{10^{-6} \, s^{-1}}$) &  & (mK)  \\
\hline \\ [-1ex ] 
HCP &  J=1-0 & 1.9 & 39.95190 & 0.04 & 3 & 23 & 1 \\

PN & J=1-0 & 2.3  & 46.99028 & 3.04 & 3 & 214 & 2 \\

CP  & N= 1-0, J=3/2-1/2, F=2-1 &2.3  & 47.98288 &  0.43 & 5 & 51 & 3\\

PO & J=3/2-1/2, $\mathrm{\Omega}$=1/2, F= 2-1, e & 3.2 & 65.31224 & 3.83 & 5 & 12 & 4\\
\hline
\end{tabular} 
\tablebib{(1) \cite{bizzocchi}; (2) \cite{cazzoli}; (3) \cite{saito}; (4) \cite{bailleux}.}
\end{table*}

\section{Conclusions} \label{outlook}
The aim of this work is to understand through observations and chemical simulations which physical conditions favour the production of P-bearing molecules in the diffuse interstellar medium and to what degree. Observing diffuse clouds offers us the opportunity to constrain  an important parameter in our chemical simulations, which is the depletion level of phosphorus (and in general the initial elemental abundances). 

We performed single-pointing observations (IRAM 30m telescope) of the (2-1) transitions of the species PN, PO, HCP and CP at 3 mm towards the line of sight to the bright continuum source B0355+508.  None of the above transitions was detected. Nevertheless, the sensitive observations yielding an rms level  of $\sim6$ mK, have allowed us to obtain reliable upper limits (see Tables \ref{tab:upper_limits_PN_PO} and \ref{tab:comparison_prediction_upper_limits}).

We have obtained high SNR detections of the (1-0) lines of HNC, CN and $\mathrm{^{13}CO}$ between 80 and 110 GHz.  We also show a first detection of $\mathrm{C^{34}S}$ (2-1) at 96 GHz towards the two densest cloud components at $-10 \, \mathrm{km \, s^{-1}}$ and  $-17 \, \mathrm{km \, s^{-1}}$. Following this, we were able to derive a sulfur isotopic ratio $\mathrm{^{32}S/^{34}S}$ ratio of $12.8\pm4.8$ and  $18.7\pm9.5$ towards the $-10 \,\mathrm{km \, s^{-1}}$ and $-17 \,\mathrm{km \, s^{-1}}$  features, with the latter being close to the local interstellar value of $24\pm5$ \citep{chin1996}. The detected molecular species show the highest abundances towards the two components at $-10 \,\mathrm{km \, s^{-1}}$ and $-17 \,\mathrm{km \, s^{-1}}$, as already shown in previous work \citep[e.g.][and references therein]{liszt18}. 

Based on the detected molecular abundances, we updated our chemical model in order to provide reliable predictions of abundances and line intensities of P-containing molecules that will serve as a guide for future observations. For this purpose we ran a grid of chemical models, with typical physical conditions of diffuse/translucent clouds, trying to reproduce the observed abundances and upper limits of HNC, CN, CO and CS  in every cloud component along the line of sight (at $ -4,\, -8, \, -10, \, -14 \, \mathrm{and} \, -17 \, \mathrm{km \, s^{-1}}$). For the clouds with $\varv_{\mathrm{LSR}} = -10 \, \mathrm{km \, s^{-1}}$ and $-17 \, \mathrm{km \, s^{-1}}$, the best agreement between observed and modeled abundances is reached at a time $t_{\mathrm{best}}=6.2 \times 10^6$ yrs and at  $r_{\mathrm{best}}= (n\mathrm{(H)}, A_V, T_{\mathrm{gas}}) = (300 \, \mathrm{cm^{-3}}, \, 3 \, \mathrm{mag}, \, 40 \, \mathrm{K})$. We chose this set of parameters as a reference for modeling the phosphorus chemistry.

According to our best-fit model mentioned above, the most abundant P-bearing species are HCP and CP ($\sim 10^{-10}$) at a time of $t=10^7$ yrs.  The species PN, PO and $\mathrm{PH_3}$ also show relatively high predicted abundances of $4.8\times 10^{-11}$ to $1.4\times 10^{-11}$ at the end of our simulations. All species are effectively destroyed through reactions with $\mathrm{C^+}$, $\mathrm{H^+}$ and $\mathrm{He^+}$. The molecules HCP, CP and PO are efficiently formed throughout the entire chemical evolution via the dissociative electron recombination of the protonated species $\mathrm{PCH_2^+}$ and $\mathrm{HPO^+}$, respectively.  In addition, the species $\mathrm{PH_3}$ is mainly formed on dust grains through subsequent hydrogenation reactions of P, PH and $\mathrm{PH_2}$ and then released to the gas-phase via photodesorption. Finally, PN is formed at late times ($10^5-10^7$ yrs) mainly through the reaction $\mathrm{N + CP \rightarrow PN +C}$. 

We have also examined how the visual extinction $A_V$, the cosmic-ray ionisation rate $\zeta(\mathrm{CR})$ and the surface mobility on dust grains are affecting the P-bearing chemistry. We found that all P-bearing species are strongly sensitive to the visual extinction: low $A_V$ values of 1 and 2 mag lead to very low P-bearing molecular abundances of $\sim 10^{-14}-10^{-12}$, indicating that a translucent region rather than a diffuse one is needed to produce observable amounts of P-containing species. All examined species in our study are influenced by the cosmic-ray ionisation rate as well. 
An increasing $\zeta(\mathrm{CR})$ enhances the abundance of $\mathrm{He^+}$, $\mathrm{H^+}$ and $\mathrm{C^+}$, which in turn are effectively destroying all P-bearing species. 
A similar conclusion was found when changing the diffusion/desorption ratio to $E_b/E_D= 0.77$ and deactivating the possibility of quantum tunneling of light species on grain surfaces. This set up increases the $\mathrm{H^+}$ abundance, which in turn efficiently reacts with and destroys PN, PO, HCP, CP and $\mathrm{PH_3}$.  Finally, we performed a study of the P-depletion level by tracing the phosphorus chemistry from a diffuse to a dense cloud with the application of a dynamical model, that varies the density, the gas and dust temperature, the cosmic-ray ionisation rate and the visual extinction with time (see Appendix \ref{depletion}). We came to the main conlusion that at high denities of $\sim 10^5 \, \mathrm{cm^{-3}}$ atomic P is strongly depleted through freeze-out on dust grains, resulting in a significant increase of the $\mathrm{gPH_3}$ abundance. The molecules  PN, PO, HCP, CP and $\mathrm{PH_3}$ are also affected by freeze-out on grains and are strongly destroyed by their reaction with $\mathrm{H_3^+}$ when reaching the dense phase at timescales of $\sim 10^6-10^7$ yrs.

Based on the predictions of our improved diffuse-cloud model, the (1-0) transitions of HCP, CP, PN and PO are expected to be detectable with estimated intensities ranging from 10 to 200 mK. A possible detection of the above species will help us constrain even further the physical and chemical properties of our model and help us understand better the yet unknown interstellar phosphorus chemistry. 

\acknowledgements{We thank the anonymous referee for his/her comments that significantly improved the present manuscript. The authors also wish to thank the IRAM Granada staff for their help during the observations. V.M.R. has received funding from the European Union's Horizon 2020 research and innovation programme under the Marie Sk\l{}odowska-Curie grant agreement No 664931. 
Work by A.V. is supported by Latvian Science Council via the project lzp-2018/1-0170. J.C. acknowledges Dr. J. C. Laas for his support with the Python programming.}

\begin{appendix}
\normalsize
\section{The depletion of Phosphorus}  \label{depletion}
The main advantage of studying the early phases of star formation is to avoid high levels of elemental depletion and thus to constrain the initial abundances used in our model to their cosmic values. This is crucial especially for phosphorus, as the small number of detections of P-bearing species in the ISM, makes the determination of the P-depletion level quite difficult. In order to obtain an approximate estimation of the expected depletion level, we apply a dynamical model with time-dependent physical conditions that allows us to follow the chemical evolution of P-bearing species from a diffuse to a dense cloud. In particular we simulate a  ``cold" stage in which a free-fall collapse takes place within $10^6$ yrs \citep{vasyunin2013, garrod2006}.  During that time the density increases from $n\mathrm{(H) = 300 \, cm^{-3}}$ to  $\mathrm{10^5 \, cm^{-3}}$ and the visual extinction rises from 1 to 40 mag. The gas temperature is decreasing from 40 to 10 K, while the dust temperature is slightly dropping from 20 to 10 K. Finally, the cosmic-ray ionisation rate is also changing from $\mathrm{1.7\times10^{-16} \, s^{-1}}$ to $\mathrm{1.3\times10^{-17} \, s^{-1}}$.  We note here that the changes in the above mentioned physical constants happen within $10^6$ yrs, while the total chemical evolution is over $10^7$ yrs. This means that between $10^6$ and $10^7$ yrs the model becomes static with the above parameters retaining the values they reached at $10^6$ yrs. That way, we simulate a long-lived collapse that provides enough time for chemical processes such as depletion to evolve.

\begin{figure*}[! h]
	\centering
 	\includegraphics[width = 1\textwidth]{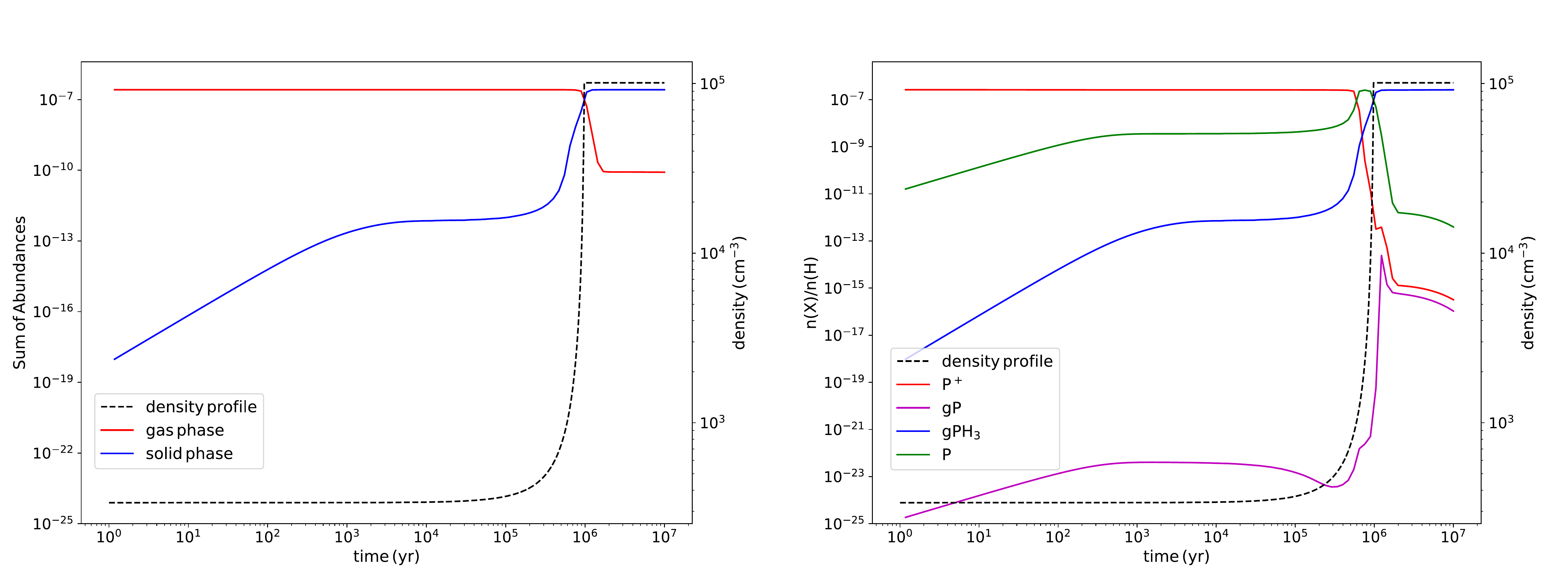}
	\caption{Results of our dynamical model that simulates the transition from a diffuse to a dense cloud. The left panel shows the sum of abundances of all P-bearing species in the gas phase (red line) and the solid phase (blue line) as a function of time. The right  panel illustrates the chemical evolution of the main carriers of phosphorus in the gas and solid phase: $\mathrm{P^+}$, $\mathrm{P}$, $\mathrm{gP}$ and $\mathrm{gPH_3}$. In both figures the density profile of the free-fall collapse is depicted as a black dashed line.}
	\label{Fig:depletion_total}
\end{figure*}

As a first step, we plot the chemical evolution of the sum of abundances of gas-phase and solid-phase P-bearing species separately (see lower left panel of Figure \ref{Fig:depletion_total}).  It is clearly visible how at late times, the gas-phase species decrease and in return, the grain species increase in abundance due to depletion. In particular, the sum of the gas-phase abundances of P-bearing species reduces by a factor of $\sim 3000$ at $t=10^7$ yrs. This does not correspond to the elemental depletion, but it indicates the redistribution of phosphorus between the gas phase and the dust grains. The right panel of Figure \ref{Fig:depletion_total}  shows the time-dependent abundances of the main carriers of phosphorus in the gas phase and on grains. The species that experience the largest change during the transition from diffuse to dense cloud are $\mathrm{P^+}$ and $\mathrm{gPH_3}$. The $\mathrm{P^+}$ abundance is strongly decreasing down to $\sim10^{-16}$, mainly through its destruction reactions with OH, $\mathrm{CH_4}$, S and $\mathrm{H_2}$.  Atomic P decreases significantly because of freeze-out on dust grains, which is also evident through the increase in gP. According to the model almost all P that freezes out, quickly reacts with hydrogen on grains and finally forms $\mathrm{gPH_3}$ (after successive hydrogenation), which reaches a high abundance of $\sim 2.5\times10^{-7}$ at the end of our simulations.

\begin{figure*}[! h]
	\centering
 	\includegraphics[width = 1.0\textwidth]{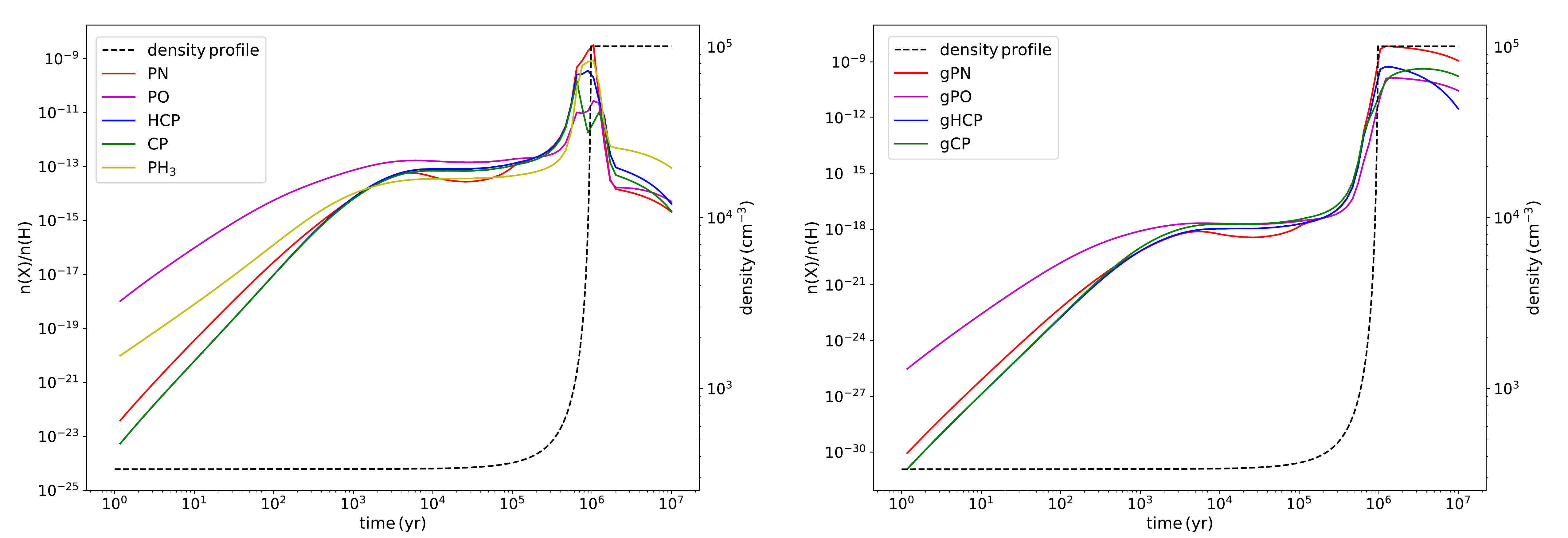}
	\caption{Chemical evolution of PN, PO, HCP, CP and $\mathrm{PH_3}$ (left panel) and the corresponding grain species (right panel) as a function of time based on our dynamical model (diffuse to dense cloud). The black dashed line illustrates the density profile of the free-fall collapse. The $\mathrm{gPH_3}$ abundance is shown in Figure \ref{Fig:depletion_total}.}
	\label{Fig:depletion_mol}
\end{figure*}

Finally, Figure \ref{Fig:depletion_mol} shows the time-dependent abundances of PN, PO, HCP, CP,  $\mathrm{PH_3}$ in the gas phase (left panel) and the corresponding grain species (right panel). All species reach their peak abundances at around $10^6$ yrs, followed by a strong decrease due to freeze-out on dust grains as well as through their reaction with $\mathrm{H_3^+}$ (at $t = 10^6-10^7$ yrs). The species PN, $\mathrm{PH_3}$ and HCP show a more significant freeze-out than CP and PO, as they are the most abundant molecules in the gas-phase at $t = 10^6$ yrs. The freeze-out process is also clearly evident from the substantial increase of the corresponding grain species once high densities of $\sim10^4-10^5 \, \mathrm{cm^{-3}}$ are reached (see right panel of Figure \ref{Fig:depletion_mol}.)

\end{appendix}

\end{document}